\documentclass[a4paper,12pt,oneside,onecolumn]{elsarticle}
\usepackage{tabularx} 
\usepackage{amsmath}  
\usepackage{amsthm}
\usepackage{graphicx} 
\usepackage[margin=1in,letterpaper]{geometry} 

\usepackage{xcolor}
\usepackage[final]{hyperref} 
\hypersetup{
	colorlinks=true,       
	linkcolor=blue,        
	citecolor=blue,        
	filecolor=magenta,     
	urlcolor=blue         
}

\usepackage{amssymb}
\usepackage{caption}
\usepackage{subcaption}
\usepackage{tikz}
\usepackage{soul}
\usepackage{slashed}
\usetikzlibrary{patterns}
\usetikzlibrary{shapes.arrows}
\usepackage{pgfplots}
\usepackage{placeins}
\pgfplotsset{compat=newest}
\usepackage{mathtools} 
\usepackage{diagbox}
\usepackage{fixltx2e}
\usetikzlibrary{arrows.meta}
\usetikzlibrary{calc,decorations.markings,fit,shapes,chains,arrows}
\usetikzlibrary{fillbetween,backgrounds}
\usetikzlibrary{patterns}
\usepackage{lineno}
\modulolinenumbers[5]
\usepackage{adjustbox}
\usepackage{booktabs}%
\usepackage{algorithm}
\usepackage{algorithmic}
\newtheorem{remark}{Remark}
\usepackage{makecell}

\begin{document}

\begin{frontmatter}

\title{A Phase-Field Formulation of Frictional Sliding Contact for 3D Fully Eulerian Fluid-Structure Interactions}

\author[ubc]{Biswajeet Rath\corref{cor1}}
\ead{biswajee@mail.ubc.ca}
\cortext[cor1]{Corresponding author}

\author[ubc]{Rajeev K. Jaiman}
\ead{rjaiman@mail.ubc.ca}

\address[ubc]{Department of Mechanical Engineering, The University of British Columbia, Vancouver, BC V6T 1Z4}

\begin{abstract}
Frictional sliding contact in hydrodynamic environments can be found in a range of engineering applications. Accurate modeling requires an integrated numerical framework capable of resolving large relative motions, multiphase interactions, and nonlinear contact responses.
Building on our previously developed fully Eulerian fluid-structure formulation, we introduce a phase-field based formulation for dynamic frictional contact in three dimensions. While the structural deformations are represented through the evolution of individual left Cauchy–Green strain tensors, contact detection is achieved via the overlap of diffuse interfaces of colliding solids.  Specifically, we utilize phase fraction functions to estimate the amount of overlap between the two interfaces, thereby circumventing explicit distance computations in 3D.
The normal contact response is defined as a volumetric body force proportional to the overlap parameter, while the tangential response is computed using the Coulomb friction model. The direction of the friction forces are derived by projecting phase-averaged relative velocities onto the local tangent plane of colliding bodies.
This proposed unified treatment enables the computation of both normal and frictional forces within a single momentum balance equation, avoiding separate velocity fields for individual solids.
We present several test cases with increasing complexity to verify and demonstrate our proposed frictional contact model. 
Verification against the Hertzian contact problem shows excellent agreement with the analytical solution, with errors below $3\%$ in the traction profile.
In the sliding block benchmark, the computed displacement profiles closely follow the analytical solution for point-mass systems across multiple friction coefficients. The ironing problem demonstrates stable force predictions under finite deformation, with normal and tangential forces matching kinetic friction laws.
The robustness and scalability of the proposed formulation are further demonstrated through a representative ship-ice interaction scenario with free-surface and frictional sliding effects.

\smallskip
\smallskip


\end{abstract}
\end{frontmatter}


\section{Introduction} \label{sec1}
Fluid-structure interaction (FSI) is a complex and widespread phenomenon observed in both natural and engineering systems. It arises from the relative motion between fluids and solids in internal and external flows, leading to forces on the solid that cause deformation and/or displacement. Multiphase FSI further complicates the problem by introducing the interaction of multiple fluid phases with several solid components. Examples of multiphase FSI are abundant in marine engineering, ranging from small-scale propeller cavitation \cite{lak2024numerical, darbhamulla2024finite, kashyap2021robust, kashyap2023unsteady, cheng2025flow} to large-scale free-surface waves generated by ship motion in oceans \cite{huang2020ship, joshi20183d}. This study is primarily motivated by ship-ice interactions in the Arctic, particularly collisions with ice floes and large ice sheets. The modeling framework is designed to capture a variety of structural interactions and interface types involving fluid-structure coupling and free-surface effects.

Simulating ship-ice interactions requires resolving drifting, large deformations, collisions, and fracture or fragmentation of ice \cite{lu2018large, li2020finite, shi2023numerical}. These complex processes pose significant challenges for computational models, particularly in describing contact mechanics accurately while also resolving hydrodynamic effects. In our previous works \cite{mao20243d, rath2023interface}, we developed a phase-field based fully Eulerian FSI framework to capture drifting and large deformations of elastic bodies within multiphase environments. In this study, we extend this framework to handle FSI-contact scenarios. Ship-ice collisions involve both normal impact and frictional sliding contact; therefore, the model must be robust enough to capture both types of interactions without compromising the accuracy of FSI predictions.

Algorithmically, contact modeling is a two-step process. The first step involves contact detection, i.e. identifying which bodies or parts of bodies are about to collide. The second step involves satisfying a set of constraints on displacement and forces on the solid bodies. Traditionally, contact problems have been primarily modeled in body-fitted Lagrangian formulations where each body is assigned its own separate computational mesh that moves with the deformation of the elastic objects. Some popular approaches include the node-to-surface \cite{hughes1976finite, wriggers1990finite, taylor1991patch} and surface-to-surface \cite{taylor1991patch, zavarise1998segment} algorithms. Both of these algorithms are based on defining a pair of master and slave bodies. Contact is said to occur when the slave body intersects/collides with the master body. Contact constraints are usually satisfied using the penalty method, Lagrange multiplier, barrier method, etc. \cite{wriggers2006computational}.

In addition to normal contact forces, the frictional response is also critical when modeling collisions between bodies. This is particularly relevant for ship-ice interactions, where sliding contact can cause severe hull damage. Although extensively studied, friction remains not fully understood, as the response depends on multiple factors such as normal pressure, relative tangential velocity, surface roughness, and temperature. In the literature, several numerical models of varying complexity have been proposed to simulate friction \cite{wriggers2006computational}. Early studies considered dry contact within the framework of plasticity \cite{bowden1964friction, michalowski1978associated}, while others developed constitutive models for friction in metal forming processes \cite{tabor1981friction}. More recent efforts have explored the micromechanical origins of friction \cite{stupkiewicz2007micromechanics}. Within the computer graphics community, friction modeling has also received attention in the context of realistic cloth and contact simulations \cite{verschoor2019efficient, larionov2021frictional}.

At the contact interface, the tangential response can be classified into two categories. The first is stick, where no relative tangential motion occurs within the contact zone. The second is slip, which involves relative tangential motion between the colliding bodies. Capturing the transition between stick and slip remains one of the central challenges in numerical friction modeling. The Coulomb law provides the simplest model of frictional contact, introducing only a single parameter, the friction coefficient. Despite its simplicity, Coulomb's law effectively captures most frictional effects in practical applications. The majority of research on frictional contact mechanics, however, has been developed in a Lagrangian description, typically using either master-slave approaches \cite{wriggers1990finite} or mortar methods \cite{mcdevitt2000mortar, puso2004mortar} for contact dynamics.
Body-fitted methods face mesh entanglement issues when dealing with translation and large deformation problems. Some of the Lagrangian contact models also face challenges when modeling new contact surfaces and capturing unbiased contact forces between the slave and master bodies. Thus, we utilize our fully Eulerian formulation to develop an integrated and versatile model capable of handling topological changes such as contact along with large deformations and displacements of the solid in the FSI-contact problem.


There has been a recent rise in Eulerian contact models for both dry and fluid-structure interaction (FSI) problems. The Eulerian approach can significantly reduce the complexity of contact detection and enforcement of contact constraints. Recent studies have demonstrated the efficiency of handling solid–solid contact in both isolated \cite{levin2011eulerian, lorez2024eulerian} and submerged environments \cite{valkov2015eulerian, frei2016eulerian} within a fully Eulerian framework. The most common strategy is to assign separate level-set or phase-field functions to each body in the domain and apply explicit contact impulses or stresses based on the distance between pairs of colliding bodies \cite{valkov2015eulerian, lorez2024eulerian}. In their recent work, \cite{lorez2025frictional} proposed a novel phase-field based Eulerian contact algorithm for quasi-static dry contact in growth processes. Their formulation solves the full set of equations for momentum balance, phase evolution, and solid deformation for each body in the domain, with contact forces providing the coupling. This is conceptually similar to computer graphics simulations \cite{levin2011eulerian, teng2016eulerian}, where contact is defined through the intersection of order parameters, following the approach of \cite{reder2021phase}. The existence of separate velocity fields in these models enables the definition of relative sliding between bodies and hence the application of a frictional force. However, such an approach is not suitable for FSI problems, where a single unified velocity field across the physical domain is required. Moreover, the computational cost increases significantly with the number of interacting bodies, limiting scalability.

In our previous work \cite{rath2024efficient}, we introduced an efficient method for modeling multibody contact among solids with identical physical properties, motivated by the need to simulate collisions between ice pieces within the multiphase domain of ship, ice, water, and air. In this study, we extend that framework by proposing a novel phase-field based contact model to handle multibody collisions among solids with different physical properties, building on our earlier work on multiphase FSI \cite{mao20243d}. Inspired by the works in \cite{reder2021phase,lorez2024eulerian}, we define an overlap parameter that quantifies the intersection of diffuse interfaces between colliding solids. An explicit contact force is then imposed, proportional to this overlap measure. The use of a unified velocity field enables consistent simulation of fluid-structure and free-surface interactions while reducing the overall computational cost. To capture sliding contact, we introduce phase-averaged velocities for each pair of colliding bodies, which provide a representative relative motion for defining tangential forces. To the best of our knowledge, this is the first work in the literature to resolve both normal and frictional contact in multiphase FSI systems within a fully Eulerian framework.

The present article is structured as follows. In Section 2, we briefly review the multiphase fully Eulerian formulation and present the governing equations deriving from conservation laws. In Section 3, we present the phase-field based normal and frictional contact formulations. Details of the numerical implementation are presented in Section 4, which emphasizes the temporal discretization and the system of matrix equations to be solved. Section 5 presents verification and demonstration of the proposed contact model across benchmark and application problems. Finally, Section 6 concludes the study and outlines directions for future work.

\section{Fully Eulerian Formulation} \label{sec2}
The fully Eulerian formulation provides a common frame of reference for both solid and fluid domains. We transform the solid equations into the Eulerian coordinate system \cite{rath2023interface} and combine them with the fluid governing equations to form the unified continuum equations. The interfaces between the different phases are represented using the phase-field model. A strain evolution equation to capture the solid deformations completes the system of equations.
Let $\Omega$ denote the domain in the reference configuration $\boldsymbol{X}$ at $t=0$ and $\Omega^t$ denote the deformed state of the domain in spatial coordinates, $\boldsymbol{x}$. The domain is divided into several subdomains $\Omega_i^t$, where each $\Omega_i^t$ can be a fluid or a solid phase. The diffuse interface parameter $\phi_i$ varies between $[-1,1]$ in the current work. We define a phase fraction function $\chi_i(\phi_i) = \frac{1+\phi_i}{2}$ that informs us if the current location $(\boldsymbol{x},t)$ is within phase $i$ or not. In this work, we restrict our attention to interactions between incompressible Newtonian fluids and incompressible Neo-Hookean solids across all test cases.

\subsection{Continuum Equations}
The unified continuum equations consisting of the mass and momentum balance equations for the physical domains can be expressed via a phase indicator $(\phi(\boldsymbol{x},t))$ based interpolation as follows
\begin{equation}\label{eq:UC}
	\begin{aligned}
		\nabla \cdot \boldsymbol{v} &= 0 \ \  &\text{on   } \Omega, \\
		\rho \left(\frac{\partial \boldsymbol{v}}{\partial t}\bigg\rvert_x + (\boldsymbol{v} \cdot \nabla)\boldsymbol{v}\right) &= \nabla \cdot \boldsymbol{\sigma} + \boldsymbol{b} \ \  &\text{on   } \Omega,
	\end{aligned}
\end{equation}
where the physical properties and fields are defined by the following interpolation rules:
\begin{equation}\label{eq:phase_int}
		\rho = \sum_{i=1}^n \chi_i(\phi_i)\rho_i, \text{   }
		  \mu = \sum_{i=1}^n \chi_i(\phi_i)\mu_i, \text{   }
        \boldsymbol{\sigma} = \sum_{i=1}^n \chi_i(\phi_i)\boldsymbol{\sigma}_i, \text{   }
        \boldsymbol{b} = \sum_{i=1}^n \chi_i(\phi_i)\boldsymbol{b}_i .
\end{equation}
In Eq.~(\ref{eq:UC}), $\boldsymbol{v}$ denotes the velocity and $\boldsymbol{b}$ denotes the body force at each spatial point $\boldsymbol{x} \in \Omega$. In Eq. (\ref{eq:phase_int}), $\rho$ represents the density, $\mu$ represents the viscosity and $n$ is the total number of phases in the domain. The Cauchy stress for a fluid phase is given by $\boldsymbol{\sigma}_i = -p\boldsymbol{I} + \mu_i(\nabla \boldsymbol{v} + (\nabla \boldsymbol{v})^T)$, whereas the stress term for a solid phase is given by $\boldsymbol{\sigma}_i = -p\boldsymbol{I} + \mu_i(\nabla \boldsymbol{v} + (\nabla \boldsymbol{v})^T) + \mu_{s,i}^L(\boldsymbol{B}_i-\boldsymbol{I})$, with $\mu_{s,i}^L$ the shear modulus and $\boldsymbol{B}_i$ the left Cauchy-Green deformation tensor.

The semi-discrete variational form of the unified continuum equations, in conjunction with the generalized-$\alpha$ time integration method \cite{chung1993time, jansen2000generalized} can be presented as follows. Let $\mathcal{S}^h$ be the space of trial solutions whose values satisfy the Dirichlet boundary conditions and $\mathcal{V}^h$ be the space of test functions whose values vanish on the Dirichlet boundary. Thus, find $[\boldsymbol{v}_h^{n+\alpha}, p_h^{n+1}] \in \mathcal{S}^h$ such that $\forall [\boldsymbol{\psi}_h, q_h] \in \mathcal{V}^h$,
\begin{equation}\label{eq:var_UC}
	\begin{aligned}
		&\int_{\Omega} \rho\left(\partial_{t} \boldsymbol{v}_{\mathrm{h}}^{\mathrm{n}+\alpha_{\mathrm{m}}}+(\boldsymbol{v}_{\mathrm{h}}^{\mathrm{n}+\alpha} \cdot \nabla) \boldsymbol{v}_{\mathrm{h}}^{\mathrm{n}+\alpha}\right) \cdot \boldsymbol{\psi}_{\mathrm{h}} \mathrm{d} \Omega+\int_{\Omega} \boldsymbol{\sigma}_{\mathrm{h}}^{\mathrm{n}+\alpha}: \nabla \boldsymbol{\psi}_{\mathrm{h}} \mathrm{d} \Omega +\int_{\Omega} q_{\mathrm{h}}\left(\nabla \cdot \boldsymbol{v}_{\mathrm{h}}^{\mathrm{n}+\alpha}\right) \mathrm{d} \Omega \\
		&+\sum_{\mathrm{e}=1}^{\mathrm{n}_{\mathrm{el}}} \int_{\Omega_{\mathrm{e}}} \frac{\tau_{\mathrm{m}}}{\rho}\left(\rho (\boldsymbol{v}_{\mathrm{h}}^{\mathrm{n}+\alpha} \cdot \nabla) \boldsymbol{\psi}_{\mathrm{h}}+\nabla q_{\mathrm{h}}\right) \cdot \boldsymbol{\mathcal{R}}_{\mathrm{m}}(\boldsymbol{v}, p) \mathrm{d} \Omega_{\mathrm{e}} +\sum_{\mathrm{e}=1}^{\mathrm{n}_{\mathrm{el}}} \int_{\Omega_{\mathrm{e}}} \nabla \cdot \boldsymbol{\psi}_{\mathrm{h}} \tau_{\mathrm{c}} \rho \mathcal{R}_{\mathrm{c}}(\boldsymbol{v}) \mathrm{d} \Omega_{\mathrm{e}} \\
		&=\int_{\Omega} \boldsymbol{b}\left(t^{\mathrm{n}+\alpha}\right) \cdot \boldsymbol{\psi}_{\mathrm{h}} \mathrm{d} \Omega+\int_{\Gamma_{\mathrm{h}}} \boldsymbol{h} \cdot \boldsymbol{\psi}_{\mathrm{h}} \mathrm{d} \Gamma ,
	\end{aligned}
\end{equation}
where $(:)$ denotes the contraction operator.
The first line contains the Galerkin terms for the combined momentum and continuity equations. The second line contains the Petrov-Galerkin stabilization terms for the continuum equations. $\boldsymbol{\mathcal{R}}_m$ and $\mathcal{R}_c$ are the element-wise residuals for the momentum and continuity equations, respectively. The stabilization parameters $\tau_m$ and $\tau_c$ \cite{shakib1991new, johnson2012numerical} are given by
\begin{equation}
	\tau_m = \left[\left(\frac{2}{\Delta t}\right)^2 + \boldsymbol{v}_h \cdot \boldsymbol{G} \boldsymbol{v}_h + C_I \bigg(\frac{\mu}{\rho}\bigg)^2 \boldsymbol{G}:\boldsymbol{G}\right]^{-1/2},  \text{   } \tau_c = \frac{1}{\mathrm{tr}(\boldsymbol{G})\tau_m} ,
\end{equation}
where $C_I$ is a constant obtained from the element-wise inverse estimate and $\boldsymbol{G}$ is the element contravariant metric tensor. $n_{el}$ and $\Omega_e$ in Eq. (\ref{eq:var_UC}) denote the total number of elements and the volume occupied by each element, respectively.

\subsection{Phase-field Model for Multiphase FSI}\label{phase-field}
We employ the phase-field model as a mathematical construct to implicitly evolve the various interfaces during the multiphase FSI simulation. The individual phase-field functions/order parameters $\phi_i(\boldsymbol{x},t)$ vary steeply within the diffuse interface and have constant values outside it. The value of $\phi_i$ is $1$ inside phase $i$ and $-1$ outside it.
The convective form of the governing Allen-Cahn equation for the evolution of the order parameter is given by,
\begin{equation}\label{eq:AC}
	\frac{\partial \phi_i}{\partial t} + \boldsymbol{v} \cdot \nabla \phi_i = -\gamma_i(t) \left ( F'(\phi_i) - \varepsilon^2 \Delta \phi_i - \beta_i(t) \sqrt{F(\phi_i)} \right ) \text{  on  } \Omega \times [0,T] ,
\end{equation}
where $\varepsilon$ is the interface thickness parameter, $F(\phi_i) = \frac{1}{4} (\phi_i^2 - 1)^2$ is the bulk energy, $\gamma_i(t)$ is the time-dependent mobility parameter for phase $i$, and $\beta_i(t) = \frac{\int_\Omega F'(\phi_i) d\Omega}{\int_\Omega \sqrt{F(\phi_i)} d\Omega}$ is the time-dependent part of the Lagrange multiplier. The time-dependent mobility parameter $\gamma_i(t)$ helps preserve the shape of the interface by minimizing the amount of convective distortion \cite{mao2021variational}, while the Lagrange multiplier term aids in global mass conservation of the order parameter field. 

The phase-field model is notorious for gradual rounding of the sharp edges when evolving solid phases. To alleviate this issue, we utilize the gradient minimizing velocity (GMV) field for convecting solid phases in the domain \cite{mao2023interface}. The purpose of this auxiliary velocity field is to extend the solid velocity into the diffuse interface so that the interface moves with the solid body, thus preserving the geometry of the solid. The governing equation for this velocity field can be formulated as
\begin{equation}
    \chi_i(\phi_i) (\boldsymbol{w}_i - \boldsymbol{v}) + (1 - \chi_i(\phi_i)) \left ((-\varepsilon \nabla \phi_i \cdot \nabla)\boldsymbol{w}_i - \frac{\varepsilon}{2\sqrt{2}}\Delta \boldsymbol{w}_i \right ) = 0,
\end{equation}
where $\boldsymbol{w}_i$ is the gradient-minimizing velocity field for the solid phase $i$. For the variational form of GMV, we define $\mathcal{S}^h$ to be the space of trial solutions whose values satisfy the Dirichlet boundary conditions and $\mathcal{V}^h$ to be the space of test functions whose values vanish on the Dirichlet boundary. We find $\boldsymbol{w}_{i,\mathrm{h}}^{n+\alpha} \in \mathcal{S}^h$ such that $\forall \boldsymbol{\psi}_h \in \mathcal{V}^h$,
\begin{equation}
\begin{aligned}
    \int_{\Omega} \chi_i(\phi_i^{\mathrm{n+\alpha}}) (\boldsymbol{w}_{i,h} - \boldsymbol{v}_{h}^{\mathrm{n+\alpha}})\cdot \boldsymbol{\psi}_h \mathrm{d}\Omega &+ \int_{\Omega} (1-\chi_i(\phi_i^{\mathrm{n+\alpha}})) ((-\varepsilon \nabla \phi_{i,h}^{\mathrm{n+\alpha}} \cdot \nabla)\boldsymbol{w}_{i,h})\cdot \boldsymbol{\psi}_h \mathrm{d}\Omega \\
    & + \int_{\Omega} (1-\chi_i(\phi_{i,h}^{\mathrm{n+\alpha}})) \frac{\varepsilon}{2\sqrt{2}} \nabla \boldsymbol{w}_{i,h} : \boldsymbol{\psi}_h \mathrm{d} \Omega = 0.
\end{aligned}
\end{equation}
The above elliptic equation for GMV is solved via Newton-Raphson sub-iterations within every nonlinear iteration. Thus, Eq. (\ref{eq:AC}) is now modified to the following form for solid phases,
\begin{equation}\label{eq:AC_GMV}
	\frac{\partial \phi_i}{\partial t} + \boldsymbol{w}_i \cdot \nabla \phi_i = -\gamma_i(t) \left ( F'(\phi_i) - \varepsilon^2 \Delta \phi_i - \beta_i(t) \sqrt{F(\phi_i)} \right ) \text{  on  } \Omega \times [0,T] ,
\end{equation}

For the variational form of the Allen-Cahn equation, we define $\mathcal{S}^h$ to be the space of trial solutions, whose values satisfy the Dirichlet boundary conditions and $\mathcal{V}^h$ to be the space of test functions whose values vanish on the Dirichlet boundary. The stabilized form for each Allen-Cahn equation can be stated as: find $\phi_{i,\mathrm{h}}^{n+\alpha} \in \mathcal{S}^h$ such that $\forall \psi_h \in \mathcal{V}^h$
\begin{equation} \label{AC_var}
	\begin{aligned}
        \int_{\Omega}( \psi_{\mathrm{h}} \partial_{\mathrm{t}} \phi_{i,\mathrm{h}}^{\mathrm{n}+\alpha_m} + \psi_{\mathrm{h}} ( \boldsymbol{v}^{\mathrm{n}+\alpha} \cdot \nabla \phi^{\mathrm{n}+\alpha}_{i,\mathrm{h}}) &+ \gamma^{\mathrm{n}+\alpha} ( \nabla \psi_{\mathrm{h}} \cdot (  \varepsilon^2 \nabla \phi_{i,\mathrm{h}}^{\mathrm{n}+\alpha} ) +\psi_{\mathrm{h}} s \phi_{i,\mathrm{h}}^{\mathrm{n}+\alpha} - \psi_{\mathrm{h}} f ) ) \mathrm{d} \Omega   \\      
		&+\sum_{\mathrm{e}=1}^{\mathrm{n}_{\mathrm{el}}} \int_{\Omega_{e}}\left(\boldsymbol{v}^{\mathrm{n+\alpha}} \cdot \nabla \psi_{\mathrm{h}}\right) \tau_{\phi} \mathcal{R}(\phi_{i,\mathrm{h}}) \mathrm{d} \Omega_{e} = 0,			
	\end{aligned}
\end{equation}
where $s$ and $f$ are the reaction coefficient and source terms, respectively, and $\mathcal{R}\left(\phi_{i,\mathrm{h}}\right)$ is the element-wise residual of each Allen-Cahn equation as defined in \cite{joshi2020variational}. The first line of Eq. (\ref{AC_var}) contains the Galerkin terms, and the second line contains the linear stabilization terms with the stabilization parameter, $\tau_{\phi}$ \cite{shakib1991new, johnson2012numerical} given by
\begin{equation}
	\tau_{\phi} = \left[\left(\frac{2}{\Delta t}\right)^2 + \boldsymbol{v} \cdot \boldsymbol{G}\boldsymbol{v} + 9\varepsilon^4 \boldsymbol{G}: \boldsymbol{G} + s^2 \right]^{-1/2} .
\end{equation}
For solid phases, $\boldsymbol{v}$ in the above equations is replaced by the gradient-minimizing velocity $\boldsymbol{w}$. The boundary and initial conditions for the Allen-Cahn equation are specified as
\begin{align}	
		\frac{\partial \phi_i}{\partial n} \biggr\rvert_{\Gamma} = \boldsymbol{n} \cdot \nabla \phi_i = 0 \ \ \ \  &\text{on   } \Gamma \times [0,T] ,\\
		\phi_i|_{t=0} = \phi_{i,0} \ \ \ \  &\text{on   } \Omega .	
\end{align}
where the initial phase-field function $\phi_{i,0}$ is defined according to the spatial distribution of phase $i$ in the domain.

\subsection{Solid Strain Evolution}
The use of a fixed computational mesh necessitates an additional transport equation to evolve the solid strain and deformation. In the current work, we capture the solid deformation by the evolution of the left Cauchy-Green deformation tensor ($\boldsymbol{B}$).
The transport equation for $\boldsymbol{B}_i$ for the solid phase $i$ moving with velocity $\boldsymbol{v}$ is given as
\begin{equation}
	\frac{\partial \boldsymbol{B}_i}{\partial t} + (\boldsymbol{v} \cdot \nabla) \boldsymbol{B}_i = (\nabla \boldsymbol{v}) \boldsymbol{B}_i + \boldsymbol{B}_i(\nabla \boldsymbol{v})^T  \ \ \ \  \text{on   } \Omega_{s,i}.
\end{equation}
Since a strain measure has no physical meaning in the fluid domain, we enforce recovery of $\boldsymbol{B}=\boldsymbol{I}$ outside the solid to prevent instabilities from propagating into the solid phase:
\begin{equation}
    \chi_i(\phi_i) \bigg( \frac{\partial \boldsymbol{B}_i}{\partial t} + (\boldsymbol{v} \cdot \nabla) \boldsymbol{B}_i - (\nabla \boldsymbol{v}) \boldsymbol{B}_i - \boldsymbol{B}_i(\nabla \boldsymbol{v})^T \bigg) + (1 - \chi_i(\phi_i))(\boldsymbol{B}_i - \boldsymbol{I}) \ \ \ \  \text{on   } \Omega.
\end{equation}
This interpolation ensures that the transport equation is solved in the solid phase and is smoothly extended into the interface.

The weak form of the transport equation for $\boldsymbol{B}_i$ is as follows. Let $\mathcal{S}^h$ be the space of trial solutions whose values satisfy the Dirichlet boundary conditions, and $\mathcal{V}^h$ be the space of test functions whose values vanish on the Dirichlet boundary. The variational form of the left Cauchy-Green tensor equation can be written as: find $\boldsymbol{B}_{i,h}^{n+\alpha} \in \mathcal{S}^h$ such that $\forall \boldsymbol{m}_h \in \mathcal{V}^h$,
\begin{equation}
	\begin{aligned}
		& \chi_i(\phi_i^{\mathrm{n+\alpha}}) \int_{\Omega} (\boldsymbol{m}_h):\bigg( \partial_t \boldsymbol{B}_{i,h}^{\mathrm{n+\alpha_m}} + (\boldsymbol{v}^{{\mathrm{n+\alpha}}} \cdot \nabla) \boldsymbol{B}_{i,h}^{\mathrm{n+\alpha}} \bigg)\mathrm{d} \Omega - \nabla \boldsymbol{v}^{{\mathrm{n+\alpha}}} \boldsymbol{B}_{i,h}^{\mathrm{n+\alpha}} - \boldsymbol{B}_{i,h}^{\mathrm{n+\alpha}} (\nabla \boldsymbol{v}^{{\mathrm{n+\alpha}}} )^T \bigg)  \mathrm{d}\Omega \\
            & + (1-\chi_i(\phi_i^{\mathrm{n+\alpha}}))\int_\Omega \boldsymbol{m}_h : (\boldsymbol{B}_{i,h}^{\mathrm{n+\alpha}} - \boldsymbol{I}) \mathrm{d}\Omega +\sum_{\mathrm{e}=1}^{\mathrm{n}_{\mathrm{el}}} \int_{\Omega_{e}} \tau_{\boldsymbol{B}}\left(\boldsymbol{v}^{\mathrm{n+\alpha}} \cdot \nabla \boldsymbol{m}_{\mathrm{h}}\right) : \boldsymbol{\mathcal{R}}(\boldsymbol{B}_{i,\mathrm{h}}) \mathrm{d} \Omega_{e} = 0,	
	\end{aligned}
\end{equation}
where $\boldsymbol{\mathcal{R}}(\boldsymbol{B}_{i,\mathrm{h}})$ denotes the element-wise residual of the strain evolution equation. The stabilization parameter is given by $\tau_{\boldsymbol{B}} = \bigg(\left( \frac{2}{\Delta t} \right)^2 + \boldsymbol{v}_h \cdot \boldsymbol{G} \boldsymbol{v}_h \bigg)^{-\frac{1}{2}}$. We obtain the updated values of $\boldsymbol{B}_i$ by solving the above variational equation in the solid phases which can then be used to calculate the new solid stresses before substituting in the unified continuum equations in the next time step. 
This completes the set of governing equations for the system. We now turn to the central theme of the paper, namely frictional contact modeling within the phase-field based FSI framework.

\section{Proposed Phase-Field Contact Formulation} \label{sec3}
In this section, we present our phase-field-based contact formulation. An overlap parameter is introduced to quantify the proximity between two colliding bodies, and the normal and tangential contact forces are formulated as functions of this parameter. We first define the overlap function and its role in detecting contact, followed by the derivation of the normal force formulation. The formulation of tangential frictional forces is then introduced using phase-averaged relative velocities.

\subsection{Overlap Parameter and Normal Contact} \label{sec1}
We define the contact force based on the overlap of diffuse interfaces between two colliding bodies. The overlap function is introduced as follows. Let us assume that we have two solids approaching for collision and represented by order parameters $\phi_1$ and $\phi_2$. We define two quantities $\chi_1$ and $\chi_2$ similar to the interpolation function for the physical properties, $\chi_1=\frac{1}{2}(1+\phi_1)$ and $\chi_2=\frac{1}{2}(1+\phi_2)$. We then define a parameter $\zeta$ that quantifies the amount of interfacial overlap between two approaching solids as follows:
\begin{equation}\label{eq:overlap}
    \zeta(\phi_1,\phi_2) = \chi_1(\phi_1) \chi_2(\phi_2).
\end{equation}
Here, $\zeta=0$ indicates no contact between the solids, while $\zeta>0$ denotes an interfacial overlap corresponding to a potential contact region.
The normal contact force, $\boldsymbol{f}_{c,n}$, is defined as a function of the overlap parameter:
\begin{equation}\label{eq:fc_n}
    \boldsymbol{f}_{c,n} = 
    \left \{
            \begin{array}{cc}
            \kappa \mu_{s,eq}^L \zeta(\phi_1,\phi_2) \boldsymbol{n}   & \mbox{if } \phi_1 \ge 0 \mbox{ or } \phi_2 \ge 0 \\
            \boldsymbol{0}   & \mbox{otherwise },
            \end{array}
    \right.
\end{equation}
where $\kappa$ is a user-defined penalty parameter controlling the magnitude of the force, and $\mu_{s,eq}^L$ is the equivalent shear modulus of the two bodies, defined as $\frac{1}{\mu_{s,eq}^L} = \frac{1-\nu^2}{\mu_{s,1}^L} + \frac{1-\nu^2}{\mu_{s,2}^L}$. The common unit normal $\boldsymbol{n}$ is given by,
\begin{equation}\label{eq:normal}
    \boldsymbol{n}=\frac{\nabla \phi_1 - \nabla \phi_2}{||\nabla \phi_1 - \nabla  \phi_2||}.
\end{equation}
The above definition of normal takes an average of the gradients of the individual order parameters at the point of interest and gives accurate directions for the normal contact force within the respective solids. 
Unlike classical penalty or Lagrange multiplier methods, where contact detection requires explicit surface discretization, the present formulation embeds the penalty directly into the diffuse-interface overlap measure, thereby avoiding surface tracking and enabling a fully Eulerian treatment of contact. In practice, $\kappa$ should be chosen sufficiently large to enforce non-penetration while maintaining numerical stability.

The contact force formulation in Eq.~\eqref{eq:fc_n} distributes the surface forces arising from collision over the overlapping volume of the two bodies and contributes directly to the volumetric body-force term in the unified momentum equation. This approach is analogous to the conventional penalty method in Lagrangian contact mechanics, but without permitting actual penetration between the bodies. The overlap function provides an elegant means of quantifying proximity and eliminates the need for explicit distance computations inherent in conventional methods. This greatly simplifies the implementation in 3D, as costly distance calculations and dynamic load balancing are avoided. Moreover, the volumetric force distribution alleviates the discontinuities often observed in Lagrangian contact forces, yielding a smooth and numerically stable contact response. This diffuse-interface formulation is particularly advantageous for large-scale multiphase FSI problems, such as ship-ice interactions, where contact regions evolve dynamically in time.

\subsection{Frictional Contact Formulation}
We now introduce the tangential component of the contact force to account for friction between colliding bodies. For simplicity, the Coulomb friction model is employed to define this tangential force under sliding contact. The magnitude of the friction force is given by
\begin{equation}
    f_{c,\tau} = C_f f_{c,n} = C_f (\kappa \mu_{s,eq}^L \zeta(\phi_1, \phi_2)),
\end{equation}
where $C_f$ is the coefficient of friction between the colliding bodies. To define the direction of friction force, we first define the tangential plane at a point given by the projection tensor $\boldsymbol{T}=(\boldsymbol{I} - \boldsymbol{n}\otimes \boldsymbol{n})$. To fix the direction, we project the relative velocity between the two bodies on the tangential plane,
\begin{equation}\label{eq:slip_rate}
    \dot{\boldsymbol{\delta}} = \boldsymbol{T}\cdot \boldsymbol{v}_{rel},
\end{equation}
where $\dot{\boldsymbol{\delta}}$ is the slip rate and $\boldsymbol{v}_{rel}$ is the relative velocity between the bodies. The unit vector $\boldsymbol{\tau}$ along the slip rate is then given by,
\begin{equation}
    \boldsymbol{\tau} = \frac{\dot{\boldsymbol{\delta}}}{||\dot{\boldsymbol{\delta}}||}.
\end{equation}
Accordingly, the frictional component of the contact force can be expressed as
\begin{equation}\label{eq:fc_t}
    \boldsymbol{f}_{c,\tau} = C_f f_{c,n} \boldsymbol{\tau}.
\end{equation}

The primary difficulty in applying this model arises from computing the relative velocity, since separate velocity fields for individual solids are not available in the Eulerian formulation. Unlike classical Lagrangian Coulomb formulations, which explicitly track surface meshes to resolve slip and stick directions, the present Eulerian formulation embeds the friction law within the diffuse-interface framework. This eliminates the need for surface tracking while still providing a consistent tangential force response. To address the absence of separate velocity fields, we introduce a novel strategy in the following subsection.

\begin{figure}[htbp]
    \centering
    \includegraphics[width=0.95\linewidth]{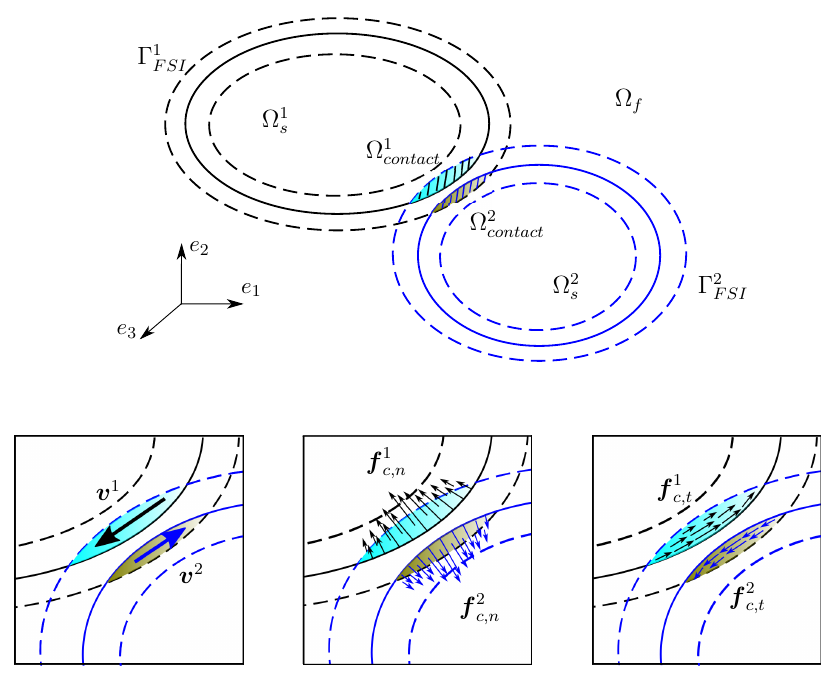}
    \caption{Schematic of two colliding bodies with diffuse interfaces (top). The shaded regions indicate the contact volumes. Insets (bottom) illustrate: phase-averaged velocities (left), normal contact forces (center), and frictional contact forces (right). Solid lines denote iso-contours of $\phi=0$, while dashed lines denote $\phi=\pm 0.9$. In the formulation, the contact volume is not restricted to the $\phi=\pm 0.9$ contours but decays smoothly into the interior; the bounded shaded regions are shown here only for clarity.}
    \label{fig:contact_schematic}
\end{figure}

\subsubsection{Computation of Relative Velocity}
In the context of contact modeling within a fully Eulerian framework, the relative velocity between solid bodies plays a crucial role in characterizing their dynamic interaction, particularly during frictional sliding. The use of a fully Eulerian model for FSI results in a unified velocity field for the whole physical domain. This is advantageous for normal contact, as the non-penetration constraint between colliding bodies is naturally enforced. However, the absence of separate velocity fields for individual solids poses a challenge for computing the slip rate required in frictional contact. Consequently, a special treatment is needed to evaluate the slip rate and consistently apply the tangential contact force within the Eulerian framework.

To compute the relative velocity between two colliding bodies, we first need to define a representative velocity for each solid. We refer to these as phase-averaged velocities. For this, we utilize the overlap volume between the diffuse interfaces of the pair of bodies. Specifically, we define two contact volumes in both bodies where the overlap parameter $\zeta$ is above a certain threshold. This describes the region where most of the contact forces are concentrated. The contact volume for body $i$ is given by the total volume of elements that satisfy $\zeta>tol$ and $\phi_i \ge 0$, where $tol$ is a user-defined threshold. This ensures that the computed velocities reflect only the regions actively engaged in contact, rather than the bulk motion of the solid.

The phase-averaged velocities for each colliding body are defined as
\begin{equation}\label{eq:v1}
    \boldsymbol{v}^1 = \frac{\int_{\Omega_{contact}^1} \boldsymbol{v} d\Omega}{\int_{\Omega_{contact}^1} d\Omega},
\end{equation}
and 
\begin{equation}\label{eq:v2}
    \boldsymbol{v}^2 = \frac{\int_{\Omega_{contact}^2} \boldsymbol{v} d\Omega}{\int_{\Omega_{contact}^2} d\Omega}.
\end{equation}
Here, $\boldsymbol{v}$ denotes the unified Eulerian velocity field over the computational domain. The relative velocity between the two sliding bodies is then expressed as
\begin{equation}\label{eq:v_rel}
    \boldsymbol{v}_{rel} = \boldsymbol{v}^1 - \boldsymbol{v}^2.
\end{equation}
These phase-averaged velocities ensure that only the active contact regions contribute to the effective velocity of each body, filtering out bulk motion away from the interface.

The sign of the friction force imposed on each body is adjusted according to the above definition of relative velocity. This approach is sufficient for our purposes, since high accuracy in the relative velocity is not required. Only its direction is needed, which is then projected onto the tangential plane to obtain the orientation along which the friction force is applied. A schematic of all contact quantities is shown in Fig.~\ref{fig:contact_schematic}, which illustrates the overlapping phase-field contours, the extracted normal direction, and the corresponding velocity vectors at the diffuse interface. This formulation enables smooth and physically consistent enforcement of contact conditions in a fully Eulerian setting, while overcoming the limitations of sharp-contact detection in traditional Lagrangian or interface-tracking methods.
\begin{remark}
It should be noted that the present framework accounts only for sliding contact and does not model sticking behavior. This assumption is justified, as the formulation is intended for hydrodynamic sliding contact problems rather than dry quasi-static solid-solid contact scenarios. Extension of the framework to capture stick-slip transitions will be considered in future work.
\end{remark}

\subsection{Contact Algorithm}

\begin{algorithm}
	\caption{3D phase-field based frictional contact algorithm}
	\label{algorithm_1}
	\begin{algorithmic}[1]
	\STATE Loop over the elements within the subgrid $\mathrm{n}_{el} = 1,2,\cdots$\\
        \STATE \quad Compute $\nabla \phi_i$ for the solid phases in the subgrid
        \STATE \quad Evaluate $\Omega_{i,contact}$ for two potentially colliding bodies \\
        \STATE \quad Compute $\boldsymbol{v}_1$ and $\boldsymbol{v}_2$ for solid phases in contact \\
    \STATE Obtain $\boldsymbol{v}_{rel}$ by subtracting the phase-averaged velocities (Eq. (\ref{eq:v_rel})) \\
    \STATE Loop over the nodes within the subgrid $\mathrm{n}_{nodes} = 1,2,\cdots$ \\
        \STATE \quad Compute common unit normal $\boldsymbol{n}$ (Eq. (\ref{eq:normal})) by transferring element gradients to nodes \\
        \STATE \quad Compute $\zeta$ using $\chi_1$ and $\chi_2$ for the pair of bodies in consideration (Eq. (\ref{eq:overlap})) \\
	    \STATE \quad Apply normal contact forces ($\boldsymbol{f}_{c,n}$) inside the solid phases ($\phi_i \ge 0$) using Eq. (\ref{eq:fc_n}) \\
		\STATE \quad Obtain the tangent plane $\boldsymbol{T}$ using the nodal normals for all points $\boldsymbol{x} \in \Omega_{i,contact}$ \\ 
        \STATE \quad Compute slip rate $\dot{\boldsymbol{\delta}}$ by projecting $\boldsymbol{v}_{rel}$ onto the tangent plane (Eq. (\ref{eq:slip_rate})) \\
		\STATE \quad Evaluate tangent vector $\boldsymbol{\tau}$ by normalizing the slip rate \\ 
        \STATE \quad Apply frictional forces ($\boldsymbol{f}_{c,\tau}$) inside the solid phases ($\phi_i \ge 0$) using Eq. (\ref{eq:fc_t})\\
    \STATE Add the volumetric contact forces to the body forces acting on the solids 
    \STATE Solve the unified momentum balance equation (Eq. \ref{eq:var_UC}) \\
	\end{algorithmic}
\end{algorithm}

The proposed contact algorithm operates within a fully Eulerian phase-field-based framework and is designed to capture contact interactions between multiple solid bodies embedded in a fixed grid.Unlike traditional Lagrangian methods, which track body surfaces explicitly, this approach relies on the overlap of diffuse interfaces represented by phase fields. Contact forces are introduced in a consistent manner using a non-linear formulation based on the degree of overlap and the relative velocity between bodies. This provides a unified treatment of normal and frictional contact in multiphase FSI.

The contact formulation described above is summarized in Algorithm~\ref{algorithm_1}. Within each nonlinear iteration, we first compute all prerequisite quantities, including common normals and relative velocities. The overlap parameter is then evaluated from the phase-field distributions in the domain. If a finite contact volume is detected in a region, the normal and tangential contact forces are computed for the nodes within that volume. The total body force at a node is then given by
\begin{equation}
    \boldsymbol{b} = \rho \boldsymbol{g} + \boldsymbol{f}_{c,n} + \boldsymbol{f}_{c,\tau},
\end{equation}
where $\boldsymbol{g}$ is the gravitational acceleration. Using this updated force, the unified momentum balance equation is solved in the computational domain to obtain the velocity and pressure fields. With the updated velocity field, the gradient-minimizing velocity $\boldsymbol{w}$ is computed and subsequently used to evolve the order parameter fields. Finally, the left Cauchy–Green deformation tensor is evolved to update the strain and stress values before proceeding to the next iteration.

The parallelized contact algorithm presented in Algorithm \ref{algorithm_1} allows us to handle frictional sliding scenarios for FSI applications. It obviates the need to solve separate momentum balance equations for each solid phase by introducing phase-averaged velocities to compute the slip rate. This reduces the computational cost significantly, since the Navier-Stokes equation is the most expensive part of the whole solution procedure. The developed framework is not restricted to quasi-static dry contact scenarios, but is capable of solving fully coupled dynamic multiphase FSI simulations.
This contact algorithm achieves robust enforcement of collision constraints without the need for surface tracking, contact meshes, or complex Lagrange multipliers. It is particularly well suited for fluid-structure interaction problems involving topological changes, fragmentation, or multibody reconfiguration, as encountered in ice floe dynamics, biomechanics, and soft robotics.

\section{Numerical Framework and Implementation Details} \label{sec5}
In this section, we present the details of the numerical implementation of the proposed framework within our in-house multiphysics/multiphase solver \cite{mao2023interface, rath2023interface}. We first elaborate on the temporal discretization and linearized system of equations. Next, we discuss some implementation details of the contact quantities in the parallelized 3D framework.

\subsection{Temporal Discretization}
The temporal discretization is carried out via the generalized-alpha time integration method \cite{chung1993time, jansen2000generalized}. This scheme introduces a single user-controlled parameter, the spectral radius $\rho_{\infty}$, which damps undesirable high-frequency oscillations in the solution. For any generic variable $\varphi$, the generalized-$\alpha$ method is expressed as
\begin{align}
	\varphi^{n+1}&=\varphi^{n}+\Delta t \partial_t \varphi^n +\Delta t\varsigma (\partial_t \varphi^{n+1}-\partial_t \varphi^n),\\
	\partial_t \varphi^{n+\alpha_m} &= \partial_t \varphi^{n}+\alpha_m (\partial_t \varphi^{n+1}-\partial_t \varphi^n),\\
	\varphi^{n+\alpha}&=\varphi^n+\alpha(\varphi^{n+1}-\varphi^{n}),
\end{align}
where $\Delta t$ is the time step size, $\alpha_m$, $\alpha$ and $\varsigma$ are the generalized-$\alpha$ parameters defined as:
\begin{align}
	\alpha=\frac{1}{1+\rho_{\infty}},\ \alpha_m=\frac{1}{2}\left(\frac{3-\rho_{\infty}}{1+\rho_{\infty}}\right),\ \varsigma=\frac{1}{2}+\alpha_m-\alpha.
\end{align} 
The method follows a predictor–multi-corrector strategy between time steps $t^{n}$ and $t^{n+1}$, with intermediate updates at $t^{n+\alpha}$. Setting $\rho_{\infty}=1$ recovers the trapezoidal time integration scheme.

\subsection{Block Matrix Equations}
The system of equations are solved in a partitioned-block iterative manner. We employ the Newton-Raphson method for the root finding process of each block. We solve for the increments of the primitive variables including velocity $(\Delta \boldsymbol{v})$, pressure $(\Delta p)$, order parameters for each phase $(\Delta \phi_i)$, the strain tensor for each solid phase $(\Delta \boldsymbol{B}_i)$ and the GMV for each solid phase $(\Delta \boldsymbol{w}_i)$. The linearized system for the unified continuum equations can be formulated as:
\begin{align} \label{LS_UCeq}
	\begin{bmatrix}
		\boldsymbol{K}_\Omega &  & \boldsymbol{G}_\Omega \\
		& \\
		-\boldsymbol{G}^T_\Omega &  &\boldsymbol{C}_\Omega
	\end{bmatrix} 
	\begin{Bmatrix}
		\Delta \boldsymbol{v}^\mathrm{n+\alpha} \\
		\\
		\Delta p^\mathrm{n+1}
	\end{Bmatrix}
	= \begin{Bmatrix} 
		-\widetilde{\boldsymbol{\mathcal{R}}}_\mathrm{m}(\boldsymbol{v},p) \\
		\\
		-\widetilde{\mathcal{R}}_\mathrm{c}(\boldsymbol{v})
	\end{Bmatrix} ,
\end{align}
where $\boldsymbol{K}_\Omega$ is the stiffness matrix of the unified momentum equation consisting of inertia, convection, diffusion and stabilization terms, $\boldsymbol{G}_\Omega$ is the discrete gradient operator, $\boldsymbol{G}^T_\Omega$ is the divergence operator and $\boldsymbol{C}_\Omega$ is the pressure-pressure stabilization term. $\widetilde{\boldsymbol{\mathcal{R}}}_\mathrm{m}(\boldsymbol{v},p)$ and $\widetilde{\mathcal{R}}_\mathrm{c}(\boldsymbol{v})$ represent the weighted residuals of the stabilized momentum and continuity equations, respectively. The updated velocity is then used to compute the GMV for each solid body. The block matrix system for the elliptic equation for GMV is given as,
\begin{align} \label{LS_GMV}
	\begin{bmatrix}
		\boldsymbol{K}_{\boldsymbol{w}}
	\end{bmatrix} 
	\begin{Bmatrix}
		\Delta \boldsymbol{w}_i^\mathrm{n+\alpha}
	\end{Bmatrix}
	= \begin{Bmatrix} 
		-\widetilde{\mathcal{R}}(\boldsymbol{w}_i)
	\end{Bmatrix} ,
\end{align}
where $\boldsymbol{K}_{\boldsymbol{w}}$ is the stiffness matrix and $\widetilde{\mathcal{R}}(\boldsymbol{w}_i)$ represents the weighted residual of the GMV equation.

The linearized form of the Allen-Cahn equation can be expressed as
\begin{align} \label{LS_AC}
	\begin{bmatrix}
		\boldsymbol{K}_{AC}
	\end{bmatrix} 
	\begin{Bmatrix}
		\Delta \phi_i^\mathrm{n+\alpha}
	\end{Bmatrix}
	= \begin{Bmatrix} 
		-\widetilde{\mathcal{R}}(\phi_i)
	\end{Bmatrix} ,
\end{align}
where $\boldsymbol{K}_{AC}$ includes the inertia, convection, diffusion, reaction and stabilization terms and $\widetilde{\mathcal{R}}(\phi_i)$ represents the weighted residual for the stabilized conservative Allen-Cahn equation.
Similarly, the linearized form for transport of the left Cauchy-Green tensor is given by:
\begin{align} \label{LS_CGT}
	\begin{bmatrix}
		\boldsymbol{K}_{CGT}
	\end{bmatrix} 
	\begin{Bmatrix}
		\Delta \boldsymbol{B}_i^\mathrm{n+\alpha}
	\end{Bmatrix}
	= \begin{Bmatrix} 
		-\widetilde{\mathcal{R}}(\boldsymbol{B}_i)
	\end{Bmatrix} ,
\end{align}
 where $\boldsymbol{K}_{CGT}$ is the stiffness matrix and $\widetilde{\mathcal{R}}(\boldsymbol{B}_i)$ is the weighted residual of the left Cauchy Green tensor equation. Solid stresses are then computed from the updated deformation tensor $\boldsymbol{B}$. Nonlinear iterations are terminated when the ratio between the $L_2$-norm of the increment and the solution falls below $5 \times 10^{-4}$, or when the maximum number of iterations is reached.

\subsection{Evaluation of Contact Quantities}
To obtain the unit normals required for imposing contact forces, the order-parameter gradients of the solid phases are first computed within each element and then transferred to the nodes via the matrix assembly process. The gradients of the colliding solids are subtracted and normalized to obtain the common unit normal, evaluated for each pair of bodies in contact.

We compute the phase-averaged and relative velocities within each subgrid/partition in a parallelized implementation. For two-body collisions, we compute the global relative velocity using MPI Reduce over the subgrids. We first obtain the total contact volume and the integral of the velocity field within the contact volume separately. Performing reduction operations on these two quantities and dividing them gives us the global phase-averaged velocities. Subsequently, subtracting them gives us the relative sliding velocity between the colliding solids. This global relative velocity is then broadcasted to all the subgrids that can be used during the local contact force application. 

For collisions involving multiple bodies of a particular phase, such as ship-ice floe collisions, we use the subgrid-level relative velocities, under a reasonable assumption of one ice floe in a subgrid. These are then used directly during the computation of contact forces in the respective subgrid.

\begin{remark}
This approach has the limitation of producing inaccurate forces when two colliding bodies lie in neighboring subgrids, since the relative velocities cannot be computed reliably. One possible remedy is to implement body-tracking tags for solids of the same phase within the domain. Such tags would allow phase-averaged velocities to be defined for each body separately at the global level, thereby improving the accuracy of relative velocity computations. For simplicity, this extension is not pursued in the present work.
\end{remark}

\section{Verification and Numerical Results}
In this section, we present verification and numerical results obtained using the proposed frictional contact algorithm. To verify the normal contact formulation, we compare our results against the analytical Hertz contact model. For verification of the friction implementation, we solve the sliding block problem and compare the displacement profiles with the analytical solutions for different friction coefficients. We then demonstrate the frictional contact formulation through the classical ironing problem, where a rigid cylinder slides over a deformable block. Finally, the developed framework is applied to a simplified ship-ice interaction scenario that incorporates free-surface effects and frictional contact dynamics.

\subsection{Hertzian Contact}\label{hertz_contact}

\begin{figure}[htbp]
    \begin{subfigure}[h]{0.36\textwidth}
        \centering
        \includegraphics[width=1\textwidth]{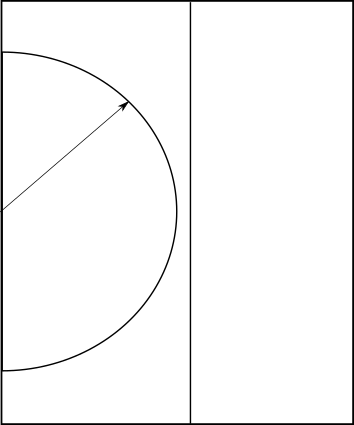}
        \put(-120,70){$\Omega_s^1$}
        \put(-50,70){$\Omega_s^2$}
        \put(-110,10){$\Omega_f$}
        \put(-120,115){$R$}
        \put(7,90){$v_x = -u_0$}
        \caption{}
        \label{hertz_schematic}
    \end{subfigure}
    \hfill
    \begin{subfigure}[h]{0.38\textwidth}
        \centering
        \includegraphics[width=1\textwidth]{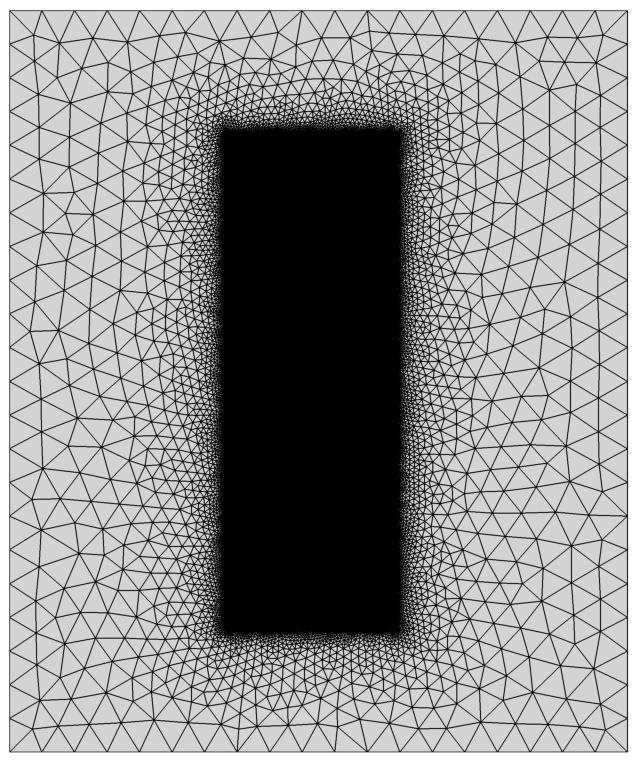}
        \caption{}
        \label{hertz_mesh}
    \end{subfigure}
    \centering
    \caption{(a) Schematic of the Hertzian contact problem with a deformable cylinder in contact with an elastic plane, and (b) single mesh layer for the same problem. The right plane wall approaches the fixed cylinder on the left with a constant velocity boundary condition of $u_0 = R/100$. The domain is extruded by 0.1 units in the out-of-plane direction to enable 3D computation. The mesh is refined in the contact zone and progressively coarsened away from it.}
    \label{fig:hertz_contact}
\end{figure}
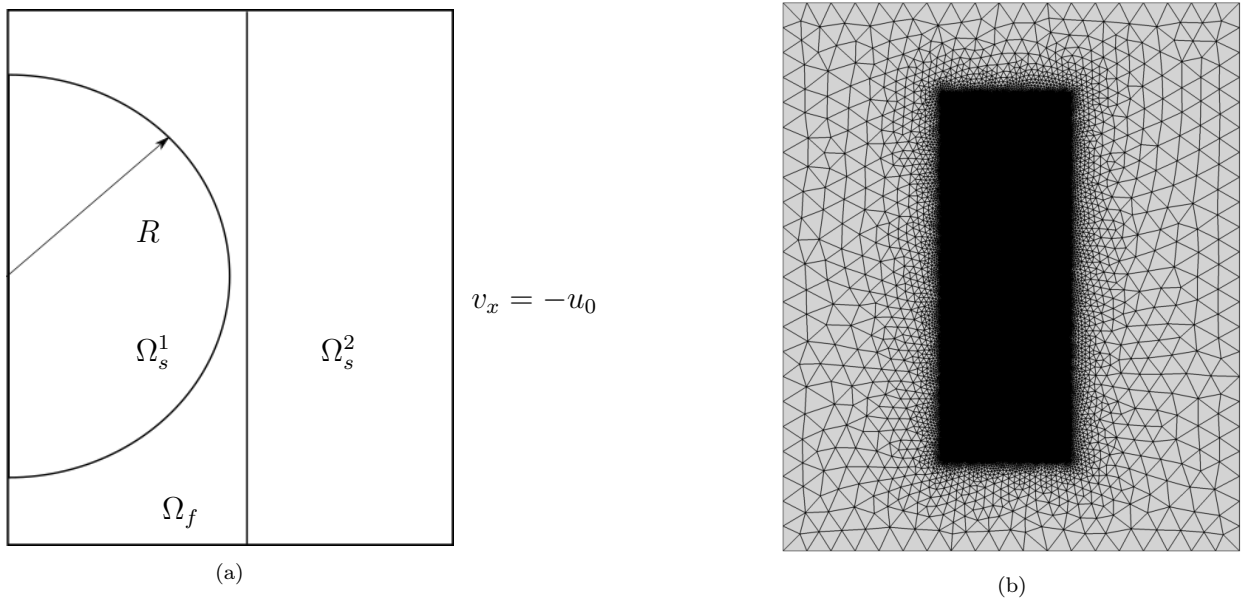

Analytical solutions for smooth Hertzian contact between two linear elastic bodies are used to assess the phase-field model in normal contact scenarios. Hertzian contact between two cylinders produces an elliptic stress distribution of width $2a$ across the contact zone. Let the cylinders have radii $R_1$ and $R_2$, Young’s moduli $E_1$ and $E_2$, and Poisson’s ratios $\nu_1$ and $\nu_2$, respectively. The half-width $a$ is given by
\begin{equation}
    a = \left( \frac{4FR_{eq}}{\pi E_{eq}} \right)^{\frac{1}{2}},
\end{equation}
where $F$ is the total contact force acting on each body, $R_{eq}$ is the equivalent radius of curvature, $R_{eq}=\frac{R_1 R_2}{R_1 + R_2}$ and $E_{eq}$ is the equivalent Young's modulus, $\frac{1}{E_{eq}} = \frac{(1-\nu_1^2)}{E_1} + \frac{(1-\nu_2^2)}{E_2}$. 
The maximum contact pressure is given by
\begin{equation}
    p_0 = \frac{2F}{\pi a}.
\end{equation}
The elliptic traction profile is given by
\begin{equation}
    \frac{p(r)}{p_0} = \left( 1 - (r/a)^2 \right)^{\frac{1}{2}} \text{       in      } 0\le r \le a.
\end{equation}
The above relations also apply to the contact of a cylinder with a plane surface, where $R=\infty$ for the plane surface. 
These relations serve as the analytical benchmark for verification of the proposed phase-field contact formulation.

\begin{figure}
    \begin{subfigure}[h]{0.36\textwidth}
        \centering
        \includegraphics[width=1\textwidth]{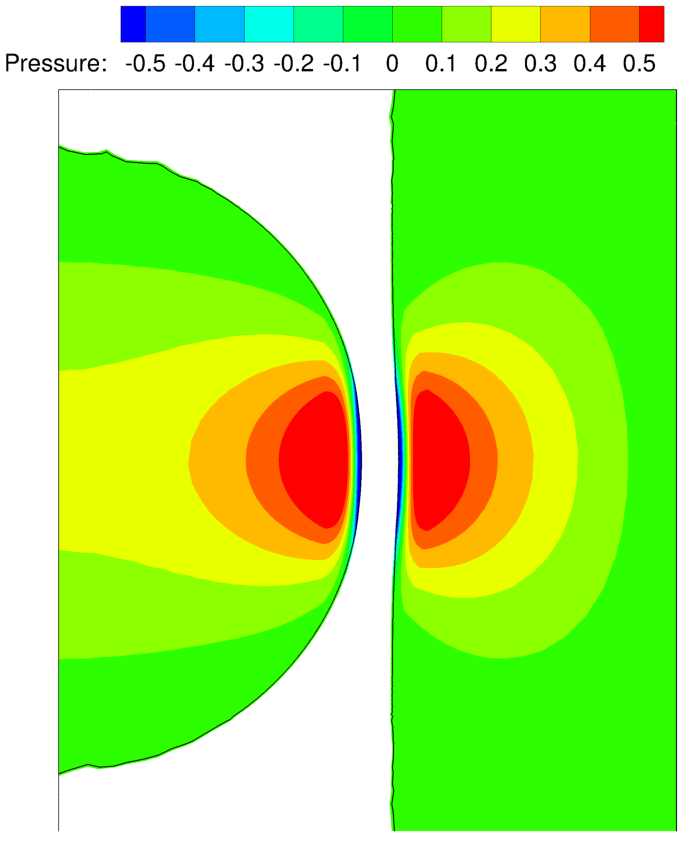}
        \caption{}
        \label{hertz_p_contour}
    \end{subfigure}
    \hfill
    \begin{subfigure}[h]{0.55\textwidth}
        \centering
        \includegraphics[width=1\textwidth]{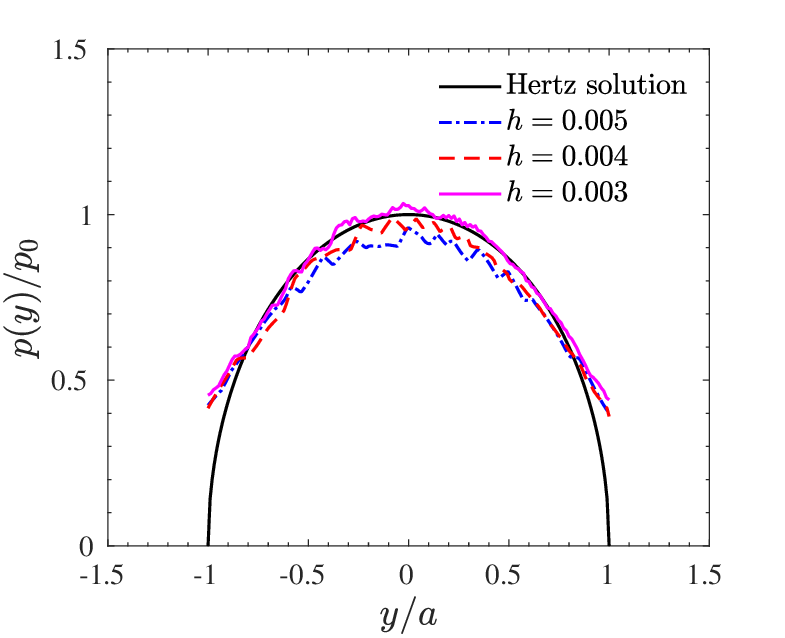}
        \caption{}
        \label{hertz_p_compare}
    \end{subfigure}
    \centering
    \caption{Hertz contact problem: (a) pressure contour for the contact problem and (b) comparison of the numerical force profiles with the analytical Hertz profile for different mesh resolutions. The interface resolution is considered to be same for all cases $\frac{\varepsilon}{h}=10$ in the contact zone.}
    \label{fig:hertz_contact}
\end{figure}

\begin{table}
\caption{\label{tab:hertz_mesh_conv} Grid convergence study for the Hertzian contact problem. $||e||_2$ is the relative $L_2$ error in the traction profile with respect to the analytical solution.}
\begin{center}
\begin{tabular}{c c c c} 
 \toprule
 \thead{Mesh} & \thead{Grid size $(h_{min})$} & \thead{No. of degrees of freedom ($N$)} & \thead{Relative error ($||e||_2$)} \\ [0.5ex] 
 \midrule
 M1 & $0.005$ & $77,112$ & $6.74\%$ \\
 M2 & $0.004$ & $138,658$ & $4.45\%$ \\ 
 M3 & $0.003$ & $242,760$ & $2.43\%$ \\ [0.5ex]
 \bottomrule
\end{tabular}
\end{center}
\end{table}

We simulate frictionless normal contact between a deformable cylinder and an elastic plane. We consider the computational domain to be $[0,2]\times [-1.2, 1.2]\times [0,0.1]$. We specify the third dimension to be one-layer thick of the mesh and set $v_z=0$ to compare with the 2D analytical solutions. The plane is assumed to be the first phase, the cylinder the second, and the background fluid the third phase. We assign very low density and viscosity to the background fluid in order to emulate dry contact. The cylinder of radius $R=1$ is centered at $(0,0)$ and clamped at the left edge, while the plane moves at a constant velocity of $v_x=R/100$ towards left, eventually pressing on the cylinder. The top and bottom faces are specified as free boundaries. The physical properties are considered to be density, $\rho_1=\rho_2=1$ and $\rho_3=0.001$, viscosity $\mu_1=\mu_2=0$ and $\mu_3=10^{-4}$ and shear modulus $\mu_{s,1}^L=\mu_{s,2}^L=10$. The contact parameters are set as $\kappa=5000$ and $C_f=0$. The simulations are run for a variety of mesh sizes, with a time-step size of $dt=0.01$. We maintain an interface resolution of $\varepsilon/h=10$ for all cases. 

The traction profile is computed by integrating the contact body force $\boldsymbol{f}_c$ normal to the interface. The total normal contact force is given by the volume integral, $F_n = \int_{\Omega_{cyl}} \boldsymbol{f}_{c,n} \cdot \boldsymbol{n} \mathrm{d}\Omega$ for the plane. The traction profile $p(y)$ along the contact patch is given by $p(y)=\int_{-a/2}^{a/2} \boldsymbol{f}_{c,n}(x,y) \cdot \boldsymbol{n} dx$. Similarly, the total tangential contact force on the plane is given by the volume integral, $F_{\tau} = \int_{\Omega_{cyl}} \boldsymbol{f}_{c,\tau} \cdot \boldsymbol{\tau} \mathrm{d}\Omega$. 
For this case, $F_{\tau} = 0$, since there is no sliding between the bodies. The computed $p(y)$ is compared against the Hertzian analytical solution for the elliptic traction profile described earlier, providing a direct benchmark for validation of the proposed contact formulation.

Figure \ref{hertz_p_contour} shows the pressure distribution over the two colliding bodies for this test case. As the solid bodies get closer to each other, the applied contact force increases in magnitude, which results in an increase in the solid pressure. This prevents the bodies from penetrating each other and causes deformations close to the interface. In Fig. \ref{hertz_p_compare}, we compare the traction profiles obtained from different meshes with the analytical Hertz solution. We can observe that the numerical profiles for the normal force over the contact patch match very well with the analytical solution. This verifies the accuracy of the proposed overlap-based contact formulation. It is worth noting that the normalization of the traction profile obscures the fact that the absolute force values are higher for smaller $\varepsilon$. Similar observations have previously been reported in the literature \cite{lorez2025frictional} and can be attributed to a decrease in the effective stiffness of the body for thicker diffuse interfaces.

The results of the grid-convergence study for this test case are reported in Table~\ref{tab:hertz_mesh_conv}. We present the convergence of the relative $L_2$ error in the traction profile for different mesh resolutions in the contact zone, denoted by $h_{min}$. For the error computation, the force profile is restricted to the region $\tfrac{p(y)}{p_0} > 0.5$ to remain within the linear regime of non-conforming contact. The relative $L_2$ error in the traction profile is defined as
\begin{equation}
    ||e||_2 = \frac{||p(y) - p_{Hertz}||_2}{||p_{Hertz}||_2},
\end{equation}
where $p_{Hertz}$ is the traction profile obtained via Hertz theory using the numerically computed total contact force $F$. We can observe that as we reduce the mesh size, the error in the traction profile decreases monotonically.

\subsection{Sliding Block Problem}
\begin{figure}[htbp]
    \begin{subfigure}[h]{0.47\textwidth}
        \centering
        \includegraphics[width=1\textwidth]{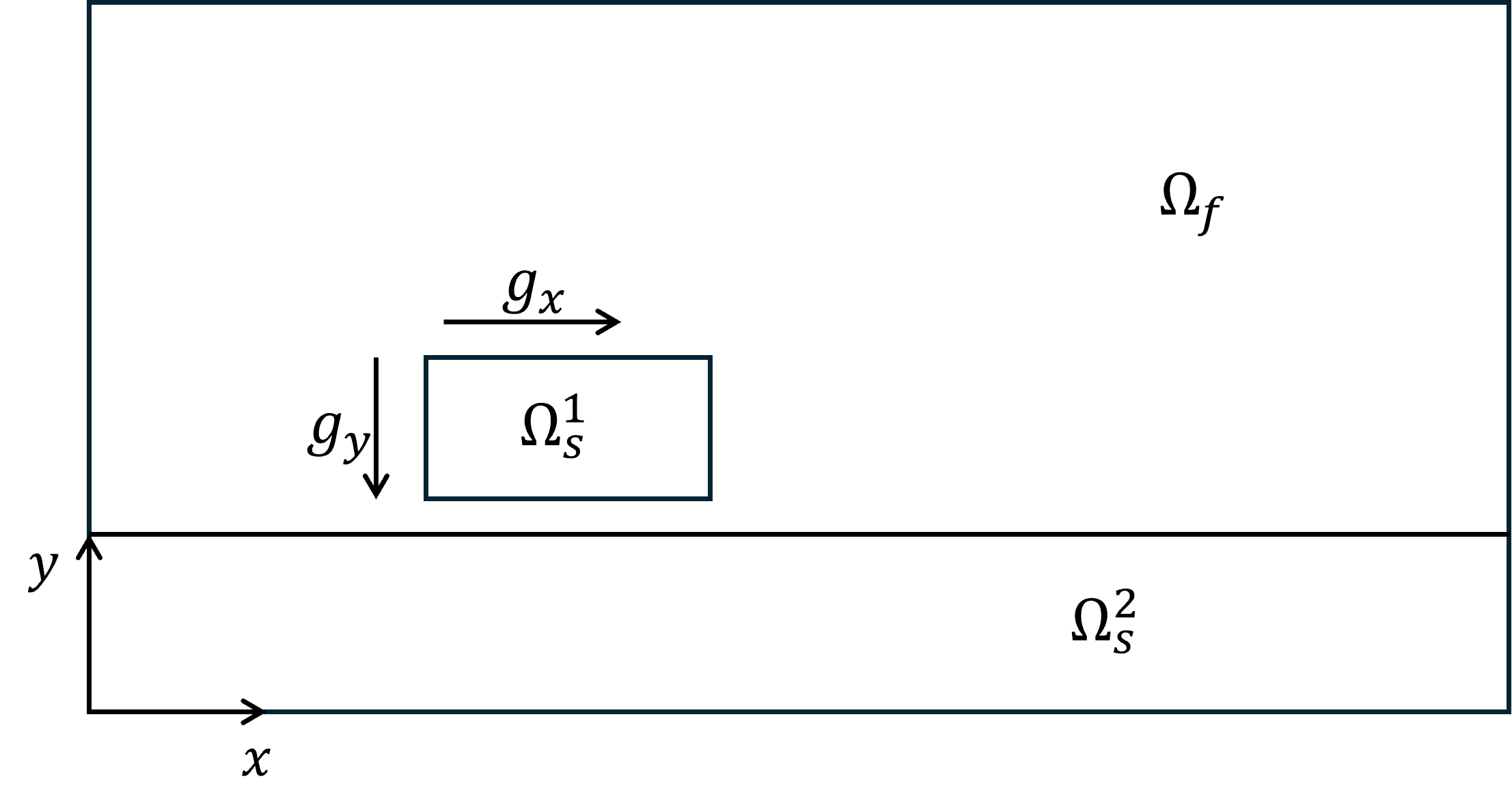}
        \caption{}
        \label{sliding_schem}
    \end{subfigure}
    \hfill
    \begin{subfigure}[h]{0.44\textwidth}
        \centering
        \includegraphics[width=1\textwidth]{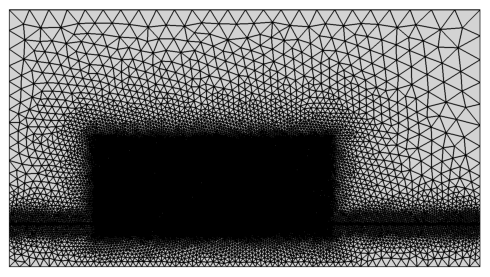}
        \caption{}
        \label{sliding_mesh}
    \end{subfigure}
    \centering
    \caption{Sliding block problem: (a) Schematic representation of the computational domain and (b) single layer of the mesh. Two components of gravity act on the elastic block, thus replicating a block sliding down an inclined plane. The domain is extruded by 0.1 units into the page to solve in 3D. The mesh is refined around the expected sliding zone and gradually coarsened elsewhere.}
    \label{fig:sliding_schem_mesh}
\end{figure}

The sliding block problem is a standard dynamic friction benchmark commonly used to verify sliding-friction implementations.
This problem has been solved by previous researchers using a variety of numerical techniques \cite{chi2015level, kamensky2019peridynamic}. Here, we simulate the sliding of an elastic block along a rigid surface using our 3D FSI-contact solver. Two components of gravity are applied to the elastic block, making the setup analogous to a block sliding down an inclined plane. For convenience, we adopt the $x$–$y$ coordinate system shown in Fig.~\ref{fig:sliding_schem_mesh} to simplify the initialization of the order-parameter fields.


The computational domain is $[0,2.2]\times[0,1.2]$ with a thickness of $0.1$ in the third dimension ($z$-axis). Two-dimensional sliding is simulated by considering a single mesh layer along the $z$-axis and enforcing $v_z=0$ throughout the domain. The rigid block occupies the region $[0,2.2]\times[0,0.2]$, while the elastic block is initialized above it within $[0.5,1.1]\times[0.27,0.47]$. The initial gap and the contact parameter are adjusted so that the vertical component of the contact force balances the body force acting on the elastic block in the vertical direction.

The elastic block is assigned as phase $\phi_1$, the rigid block as phase $\phi_2$, and the surrounding fluid as phase $\phi_3$. The material properties are: density $\rho_1=1$, $\rho_3=10^{-4}$; viscosity $\mu_1=10^{-3}$, $\mu_3=10^{-4}$; and shear modulus $\mu_{s,1}^L=1000$. The components of gravitational acceleration are set to $g_x=0.1$ and $g_y=0.1$, corresponding to an inclined plane with angle of inclination $45^\circ$. Thus, the static friction limit is $C_f=\tan(45^\circ)=1$. The contact parameter is set to $\kappa=200$, and we run cases with $C_f=\{0,0.3,0.5\}$ to ensure pure sliding. We consider a mesh of around $35,000$ nodes with a minimum element size of $h=0.007$ (Fig. \ref{sliding_mesh}). The interface thickness parameter is assumed to be $\varepsilon=0.01$. The time-step size is taken as $0.001$ and the simulation is run until $t=1$.

The bottom face is prescribed as a no-slip wall, while the side walls are treated as zero-traction boundaries. Zero-Neumann conditions are applied on all boundaries for the order-parameter fields. The bottom block is enforced to be rigid by prescribing $v_x=v_y=0$ throughout its volume. In addition, $v_y=0$ is imposed within the elastic block to restrict transverse motion and enforce pure sliding along the rigid surface. These boundary conditions ensure that the observed motion arises solely from frictional sliding.

\begin{remark}
For coarser meshes and/or thicker diffuse-interface cases, the linear interpolation of the shear modulus yields non-negligible values of $\mu_s^L$ in the narrow fluid-gap region. Nonzero values of $B_{12}$ in this region can induce spurious shear forces on the bottom face of the elastic block, potentially increasing its deceleration or even preventing motion altogether. To mitigate this effect, we enforce $\boldsymbol{B} = \boldsymbol{I}$ for strain-tensor nodes outside both blocks, i.e., $\phi_1 \le -0.95$ and $\phi_2 \le -0.95$, following approaches similar to those in \cite{sugiyama2011full, rath2024efficient}. More sophisticated interpolation strategies for the physical properties may also be explored to further minimize these artifacts.
\end{remark}

\begin{figure}
    \centering
    \includegraphics[width=0.8\linewidth]{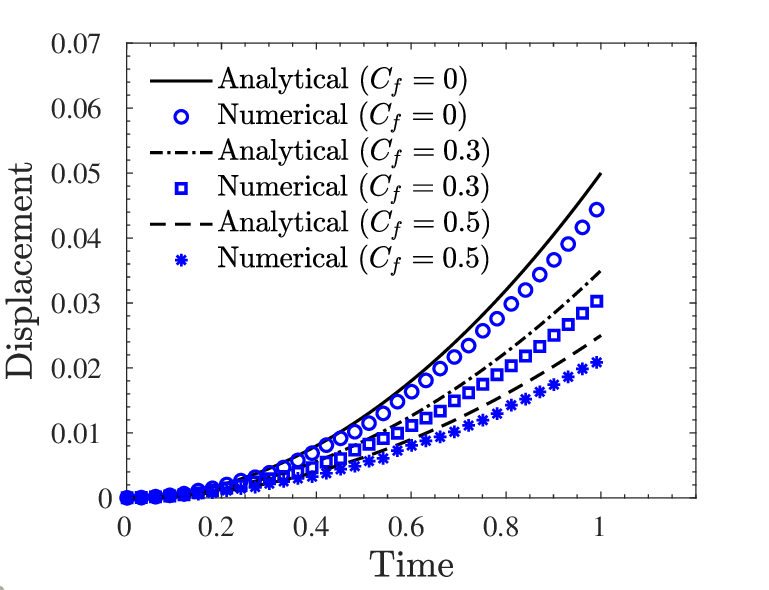}
    \caption{Sliding block problem: Comparison of the sliding displacement of the center of mass between the numerical and analytical solutions.}
    \label{fig:disp_compare}
\end{figure}

The analytical solution for the sliding displacement of the block is given by the simple kinematic equation,
\begin{equation}
    s = \frac{1}{2}a_{net}t^2,
\end{equation}
where $s$ is the displacement, $t$ is the sliding time and $a_{net}$ is the net acceleration acting on the block, given by $a_{net} = g_x - C_f g_y$.
Figure \ref{fig:disp_compare} shows the temporal evolution of the sliding displacement of the elastic block for three different values of the friction coefficient $C_f$. We can observe that the numerical solutions obtained from our 3D FSI-contact solver are in good agreement with the analytical solutions for both frictionless and frictional cases. The relative $L_2$ error in the displacement profile is between $9-14\%$ for the different cases. The deviations can be primarily attributed to the non-rigid nature of the block, with secondary contributions coming from the small fluid resistances and artifacts arising from the diffuse interface representation of the solids. The accumulation of error is due to the gradual diffusion of the interface, which increases the contact force, leading to a decrease in the net acceleration of the elastic block. We can also observe that the deviations amplify with an increase in the value of the friction coefficient $C_f$. This is due to the increase in localized friction forces on the elastic block which weakens our assumption of point mass bodies required for the analytical kinematic solutions.

\subsection{Ironing Problem}
The ironing problem is a widely used benchmark for evaluating sliding-contact formulations. It was first introduced by \cite{puso2004mortar}, who solved it using a mortar-based contact formulation with a penalty regularization scheme in a Lagrangian framework. Since then, it has been studied extensively by other researchers in the Lagrangian contact community \cite{de2011large, gitterle2012dual, temizer2013mixed}. More recently, \cite{lorez2025frictional} presented an Eulerian formulation of the frictional ironing problem for dry-contact scenarios. In the present work, we extend this problem to FSI by immersing the sliding bodies in a background fluid. To the best of our knowledge, this represents the first Eulerian FSI extension of the ironing problem.


\begin{figure}[htbp]
    \begin{subfigure}[h]{0.47\textwidth}
        \centering
        \includegraphics[width=1\textwidth]{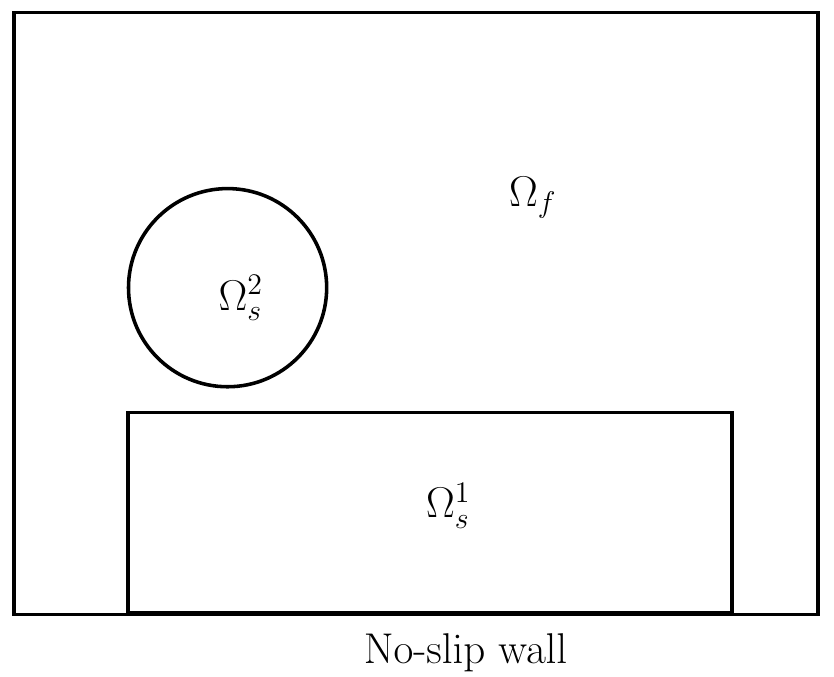}
        \caption{}
        \label{iron_schem}
    \end{subfigure}
    \hfill
    \begin{subfigure}[h]{0.47\textwidth}
        \centering
        \includegraphics[width=1\textwidth]{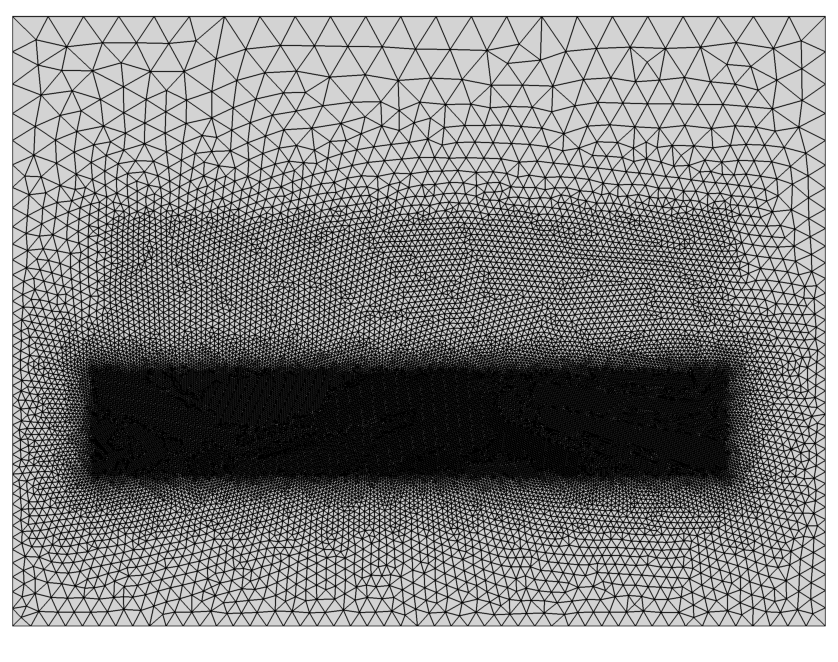}
        \caption{}
        \label{iron_mesh}
    \end{subfigure}
    \centering
    \caption{Ironing problem in fluid environment: (a) schematic of the rigid cylindrical indenter and the elastic block, and (b) single mesh layer. The indenter is displaced vertically and then translated horizontally, while the elastic block is fixed at the bottom and all other boundaries are traction-free. The domain is extruded by 0.1 units in the out-of-plane direction for 3D simulation. The mesh is refined along the sliding zone and progressively coarsened away from it.}
    \label{fig:iron_schem_mesh}
\end{figure}

In this example, we consider a computational domain of size $[0,4]\times[0,3]$ with a thickness of $0.1$ in the third dimension ($z$-axis). We simulate 2D sliding by specifying $v_z=0$ in the whole domain. A schematic of the computational domain is shown in Fig. \ref{fig:iron_schem_mesh}. We consider a disc of radius, $R=0.5$ initially centered at $[1,1.6]$. The elastic block is placed at the bottom in the region $[0.5,3.5]\times[0,1]$. This allows us to have sufficient gap and a fluid-filled region between the bodies initially. We assign the elastic block as the first phase $\phi_1$, the rigid indenter as the second phase $\phi_2$, and the surrounding fluid as the third phase $\phi_3$. The physical properties for this problem are considered to be density, $\rho_1=1$ and $\rho_3=0.001$, viscosity $\mu_1=0$ and $\mu_3=10^{-4}$ and shear modulus $\mu_{s,1}^L=1$. The contact parameters are set to $\kappa=1000$ and $C_f=0.3$.

The bottom face is prescribed as a no-slip wall, while the side walls are treated as zero-traction boundaries. Zero-Neumann conditions are applied on all boundaries for the order-parameter fields. The rigid motion of the cylinder is enforced by re-initializing its order parameter after each time step. A circular anchor zone of radius $0.3$ is defined inside the disc, and prescribed velocities are applied to the nodes within this zone. The disc is displaced vertically downward at a speed of $v_y=0.008$ for 10 time units and then translated horizontally to the right at a speed of $v_x=0.02$ for the next 90 time units.


\begin{figure}[htbp]
	\centering
	\begin{subfigure}{0.45\textwidth}
		\centering		
		\includegraphics[width=\textwidth]{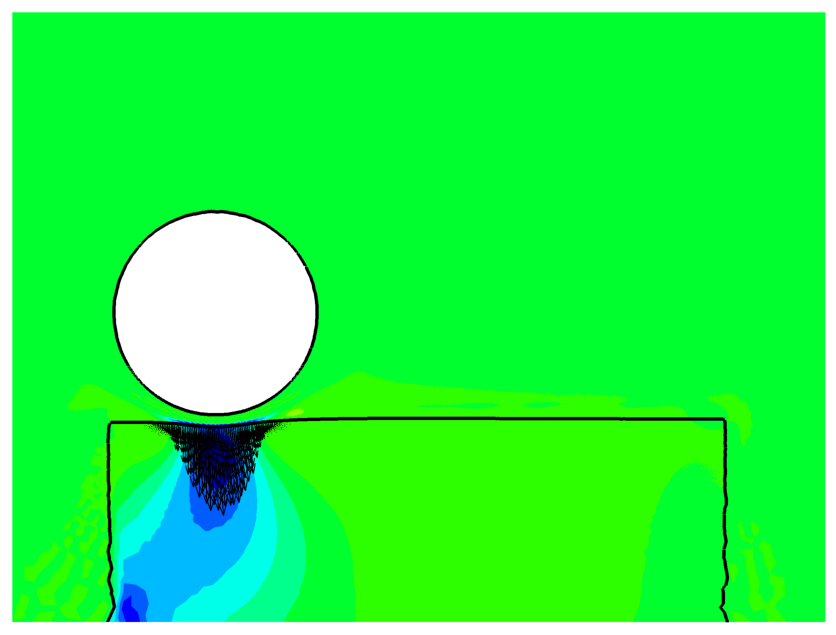}           
		\caption{}
	\end{subfigure}
	\hfill
	\begin{subfigure}{0.45\textwidth}
		\centering		
		\includegraphics[width=\textwidth]{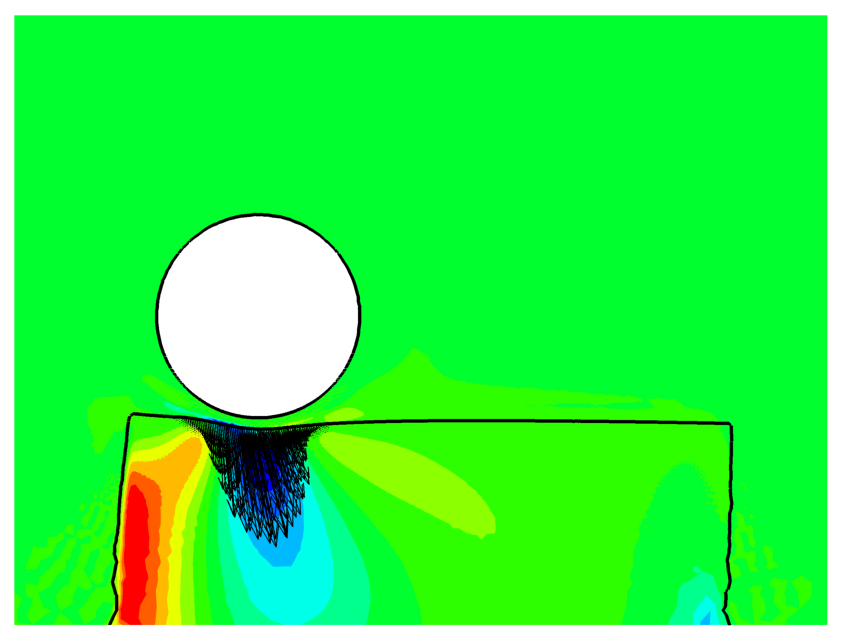}
		\caption{}
	\end{subfigure}
        \begin{subfigure}{0.45\textwidth}
		\centering		
		\includegraphics[width=\textwidth]{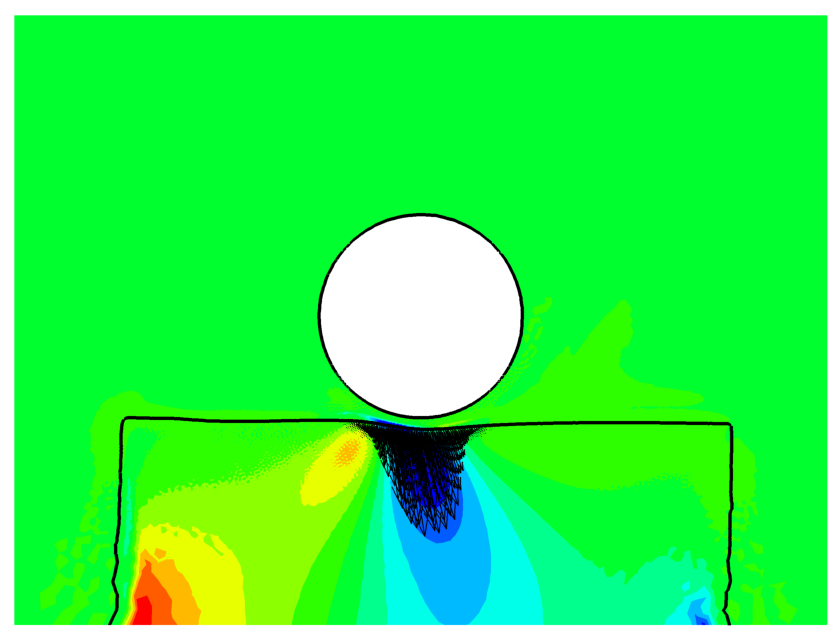}
		\caption{}
	\end{subfigure}
	\hfill
	\begin{subfigure}{0.45\textwidth}
		\centering		
		\includegraphics[width=\textwidth]{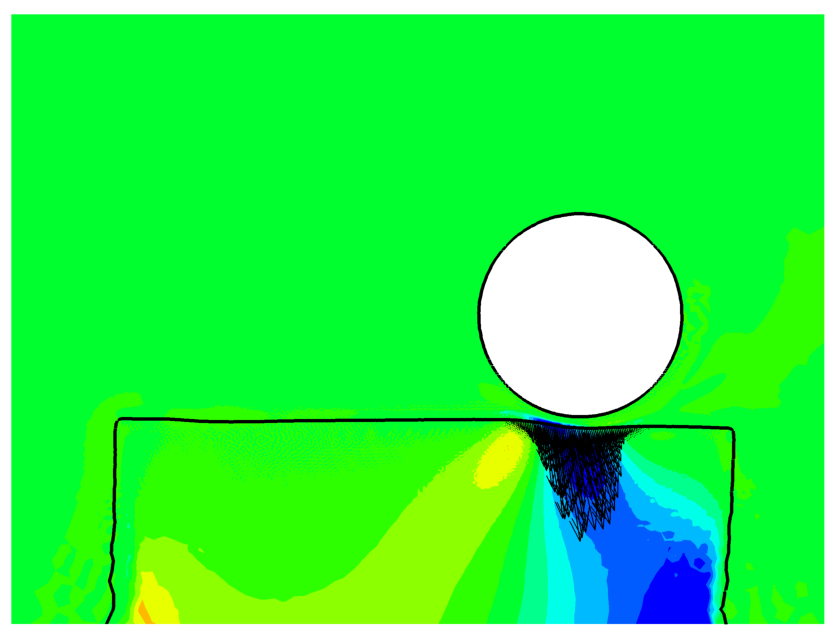}
		\caption{}
	\end{subfigure}

    \begin{subfigure}{0.4\textwidth}
		\centering		
		\includegraphics[width=\textwidth]{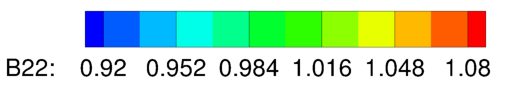}
		\label{legend_iron}
	\end{subfigure}   
	\caption{Ironing problem. Snapshots of the contours of $yy$ component of the left Cauchy-Green deformation tensor ($B_{22}$) at $t=$ (a) $10$, (b) $20$, (c) $60$ and (d) $100$. Values less than one denote regions of compression and values more than one denote regions of extension. The black arrows represent the total contact forces acting on the elastic block. The black solid lines denote the iso-contours with $\phi=0$ i.e. the boundaries of the two bodies.}
	\label{fig:iron_pltB22}
\end{figure}

\begin{figure}[htbp]
	\centering
	\begin{subfigure}{0.45\textwidth}
		\centering		
		\includegraphics[width=\textwidth]{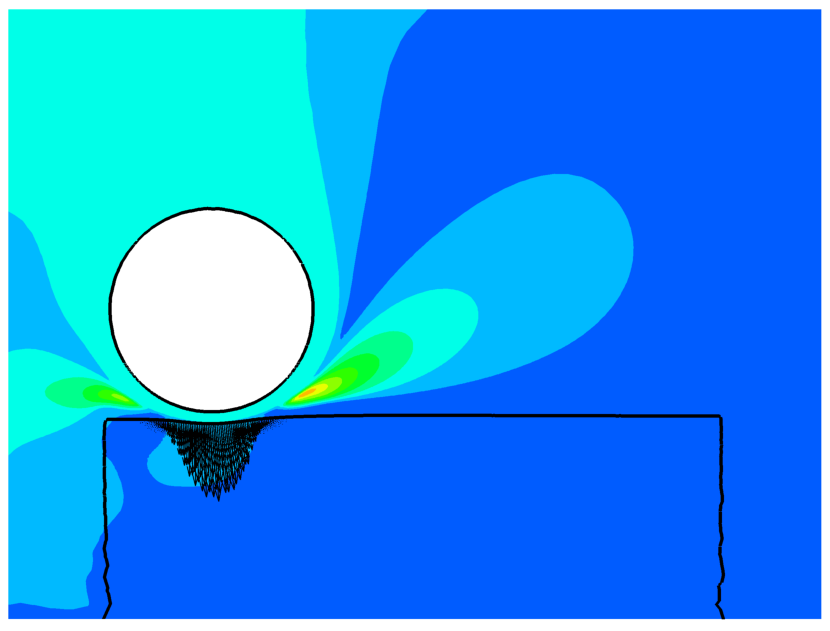}  
		\caption{}
	\end{subfigure}
	\hfill
	\begin{subfigure}{0.45\textwidth}
		\centering		
		\includegraphics[width=\textwidth]{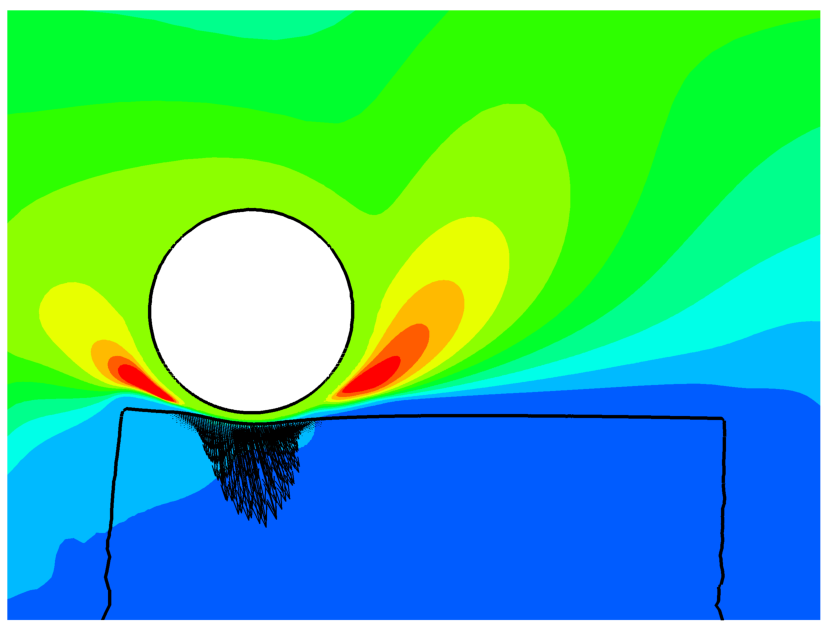}
		\caption{}
	\end{subfigure}
        \begin{subfigure}{0.45\textwidth}
		\centering		
		\includegraphics[width=\textwidth]{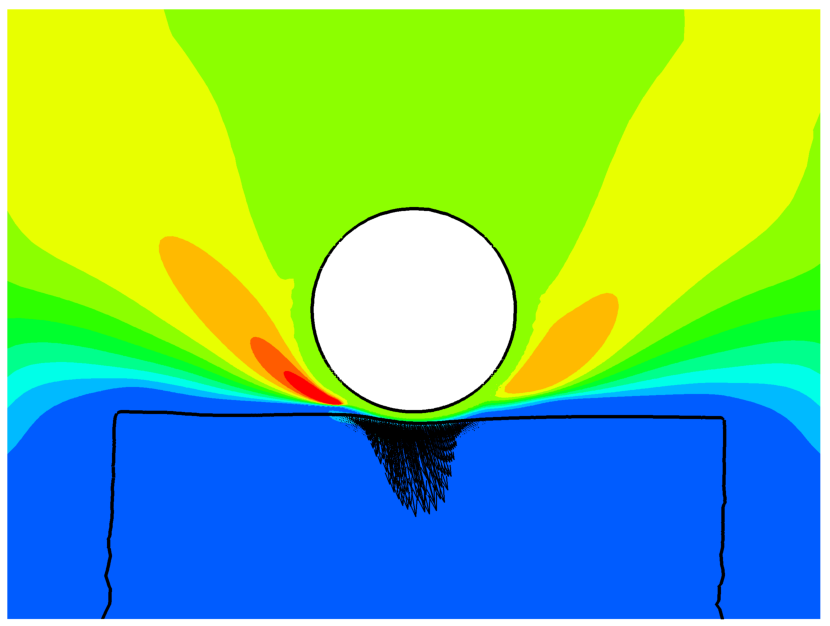}
		\caption{}
	\end{subfigure}
	\hfill
	\begin{subfigure}{0.45\textwidth}
		\centering		
		\includegraphics[width=\textwidth]{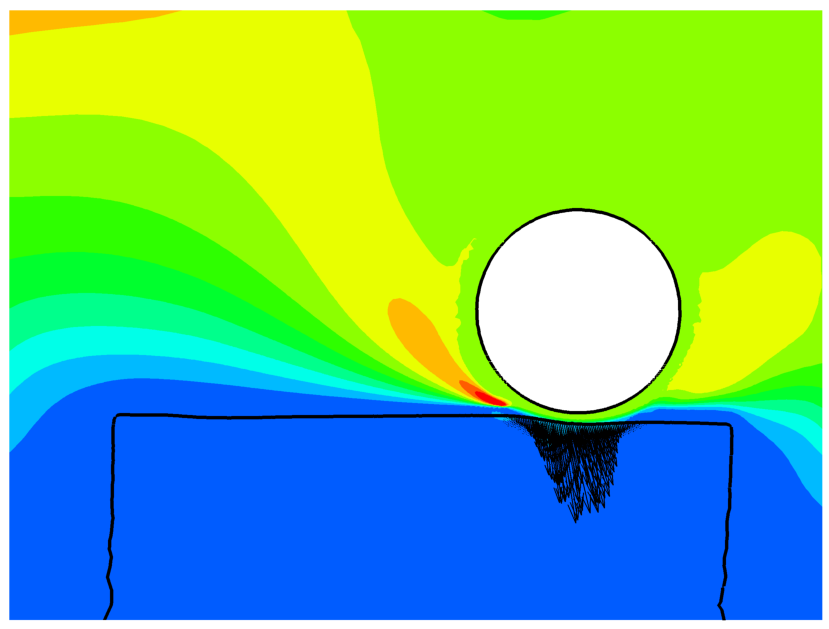}
		\caption{}
	\end{subfigure}

    \begin{subfigure}{0.4\textwidth}
		\centering		
		\includegraphics[width=\textwidth]{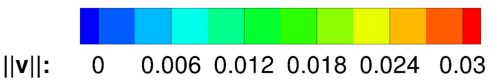}
		\label{legend_iron}
	\end{subfigure}   
	\caption{Ironing problem. Snapshots of the contours of velocity magnitude at $t=$ (a) $10$, (b) $20$, (c) $60$ and (d) $100$. The black arrows represent the total contact forces acting on the elastic block. The black solid lines denote the iso-contours with $\phi=0$ i.e. the boundaries of the two bodies.}
	\label{fig:iron_pltVMag}
\end{figure}


\begin{figure}
    \centering
    \includegraphics[width=0.6\linewidth]{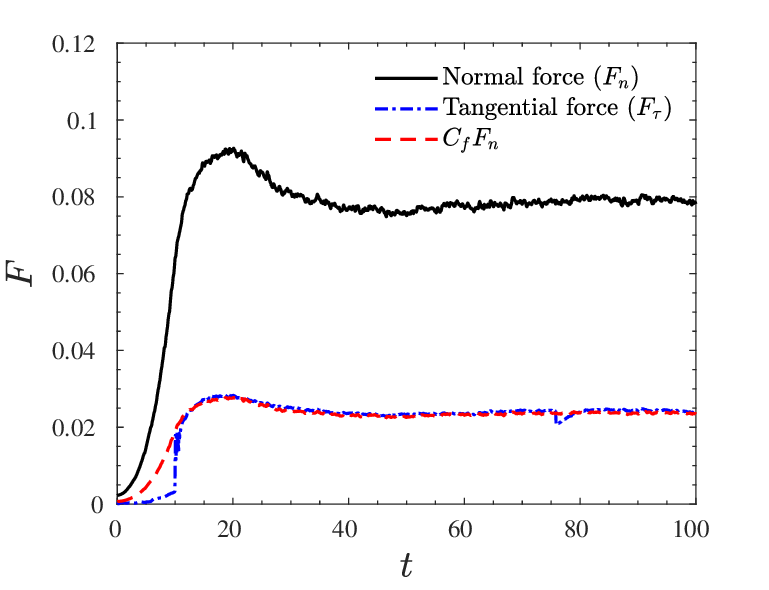}
    \caption{Ironing problem: Temporal evolution of the normal and tangential forces on the elastic block}
    \label{fig:iron_forces}
\end{figure}

Figure \ref{fig:iron_pltB22} shows the snapshots of the ironing problem at various time instances colored by the contours of the second diagonal component of the strain tensor, i.e. $B_{22}$. We can observe the compressive deformation of the elastic block directly beneath the rigid indenter at all times. The left edge of the block experiences tensile deformation as the indenter drags on the surface of the block. Similarly, the right edge of the block experiences compressive deformation as the indenter moves closer to that edge. The black arrows indicate the total contact forces acting on the elastic block. The arrows point vertically down at $t=10$, as the tangential component of the contact force is almost negligible. With rise in the frictional forces, the arrows become slightly inclined causing the block to deform accordingly. We can also observe that the arrows reach their maximum length around $t=20$, indicating the time of maximum contact forces. We explain this phenomenon later in this section. Figure \ref{fig:iron_pltVMag} illustrates snapshots with velocity magnitude contours in the domain. The higher velocities on either side of the contact zone represent the movement of the background fluid away from that region.

Figure~\ref{fig:iron_forces} shows the evolution of the normal and tangential components of the contact force as the cylinder moves across the elastic block. These forces are computed as described in section \ref{hertz_contact}. The normal force increases steadily as the cylinder is pushed downward. Notably, the peak normal force does not occur at the end of the vertical motion ($t=10$); instead, it continues to rise for some time beyond this point, attains a peak, and then stabilizes at a lower value. This phenomenon has also been reported in the literature \cite{puso2004mortar, cavalieri2015numerical, lorez2025frictional} and is attributed to rotational effects induced by the tangential frictional force, which slightly increases the depth of indentation \cite{lorez2025frictional}, thereby increasing the normal force.

The tangential force follows the line of the kinetic friction law ($C_f F_n$) closely, with only minor deviations. It rises sharply when relative sliding begins ($t>10$) and eventually stabilizes. The trend differs slightly from previous studies \cite{puso2004mortar, cavalieri2015numerical, lorez2025frictional}, primarily due to the incompressible nature of the solid assumed here. This assumption leads to immediate shape recovery after the initial compression of the block, which reduces the normal contact force acting on the elastic block. This benchmark highlights the capability of the solver to model sliding frictional contact within an FSI framework.

%
After verifying the performance of the contact model in controlled benchmark cases, we next assess its ability to handle complex, real-world geometries. In the next problem, we simulate a representative ship-ice interaction scenario involving a vessel passing through a drifting ice field composed of multiple ice floes.

\subsection{Application to Ship-ice Interaction}

\begin{figure}[htbp]
    \centering
    \includegraphics[width=0.95\linewidth]{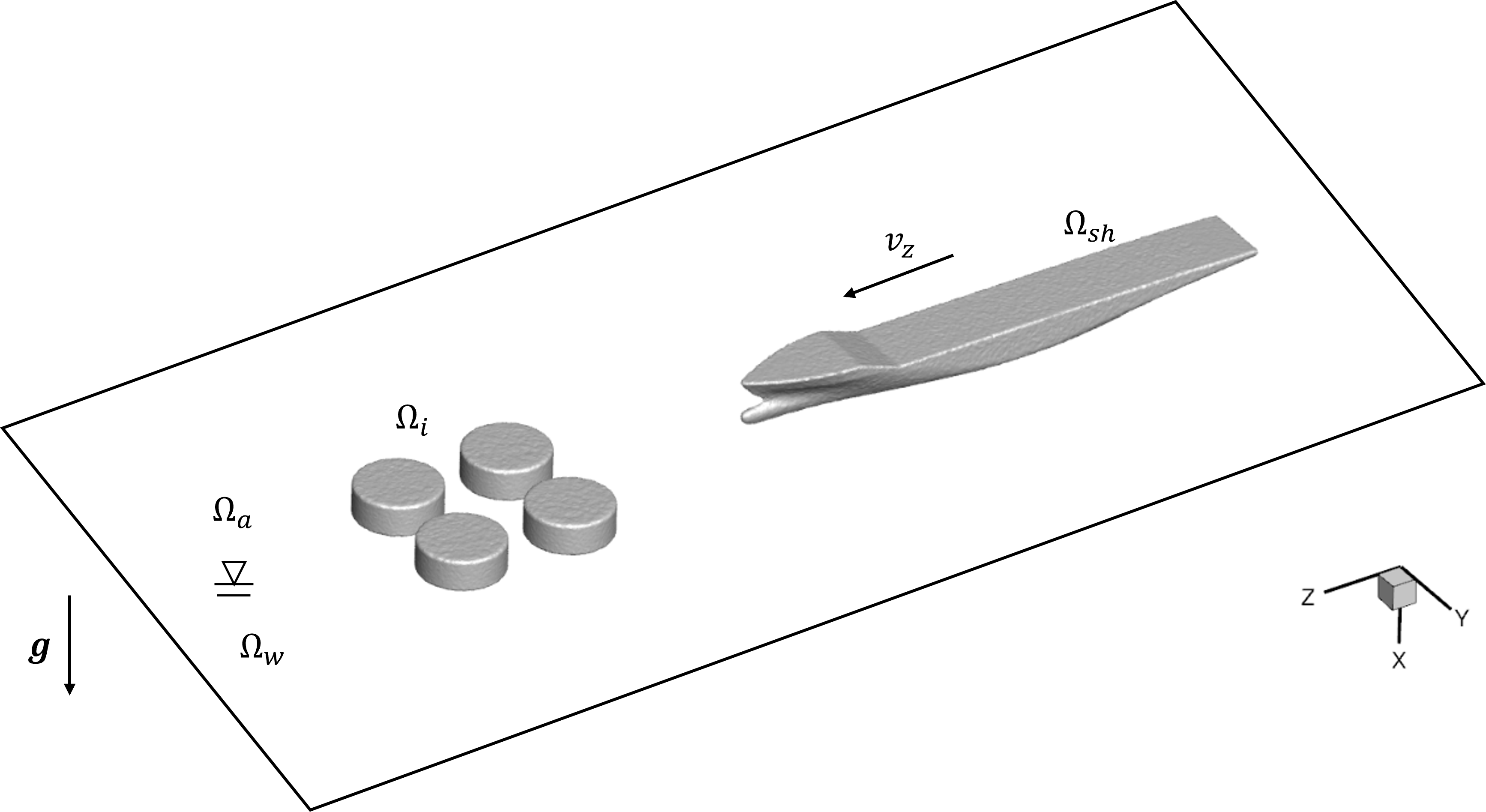}
    \caption{Isometric view of the setup for the ship-ice interaction problem.}
    \label{fig:si_schematic_isometric}
\end{figure}

To demonstrate the robustness of the developed phase-field contact model, we simulate a simplified ship-ice interaction problem with free-surface effects. The physical domain consists of four separate phases including ship, ice, water, and air, as shown in Fig. \ref{fig:si_schematic_isometric}. We assign four phase-field functions to the four phases and two strain tensors to the two solid phases, i.e. ship and ice. This setup serves as a preliminary demonstration of the framework's capability to capture multiphase FSI with contact.

\begin{figure}[htbp]
    \begin{subfigure}[h]{0.95\textwidth}
        \centering
        \includegraphics[width=1\textwidth]{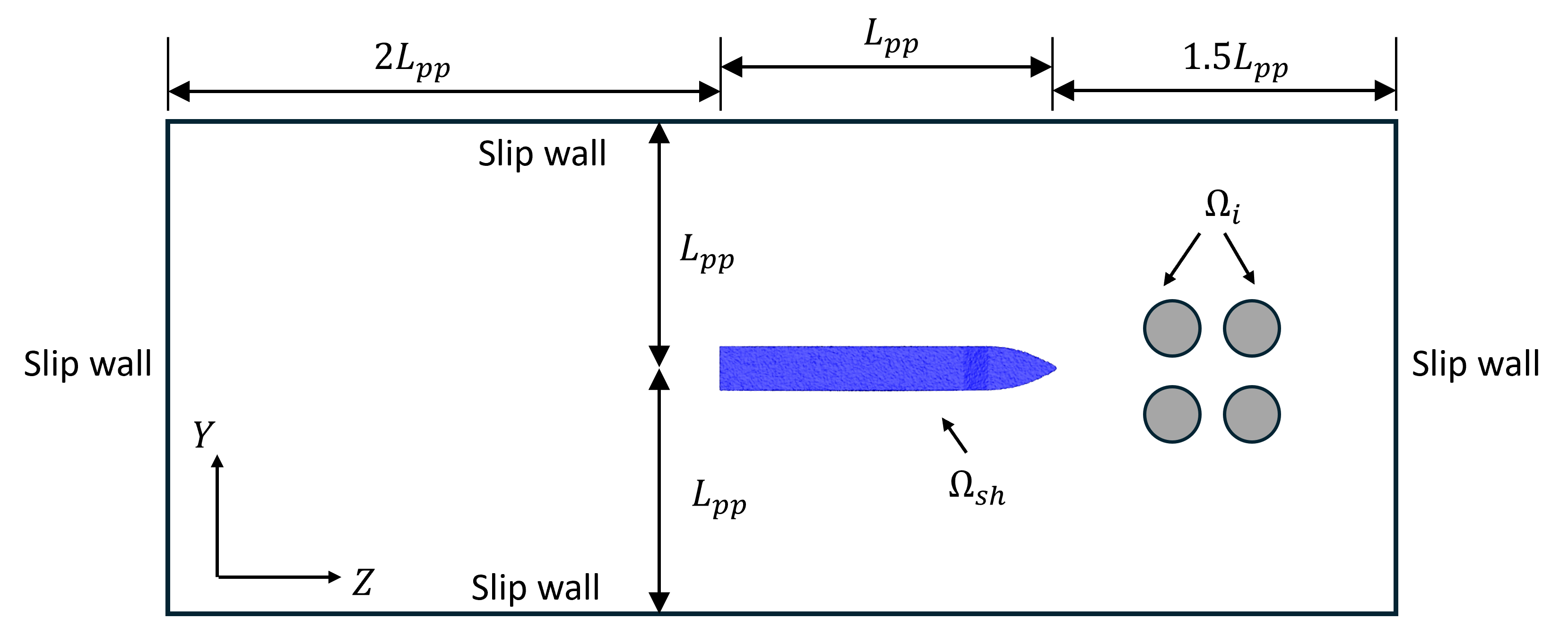}
        \caption{}
        \label{}
    \end{subfigure}
    \begin{subfigure}[h]{0.95\textwidth}
        \centering
        \includegraphics[width=1\textwidth]{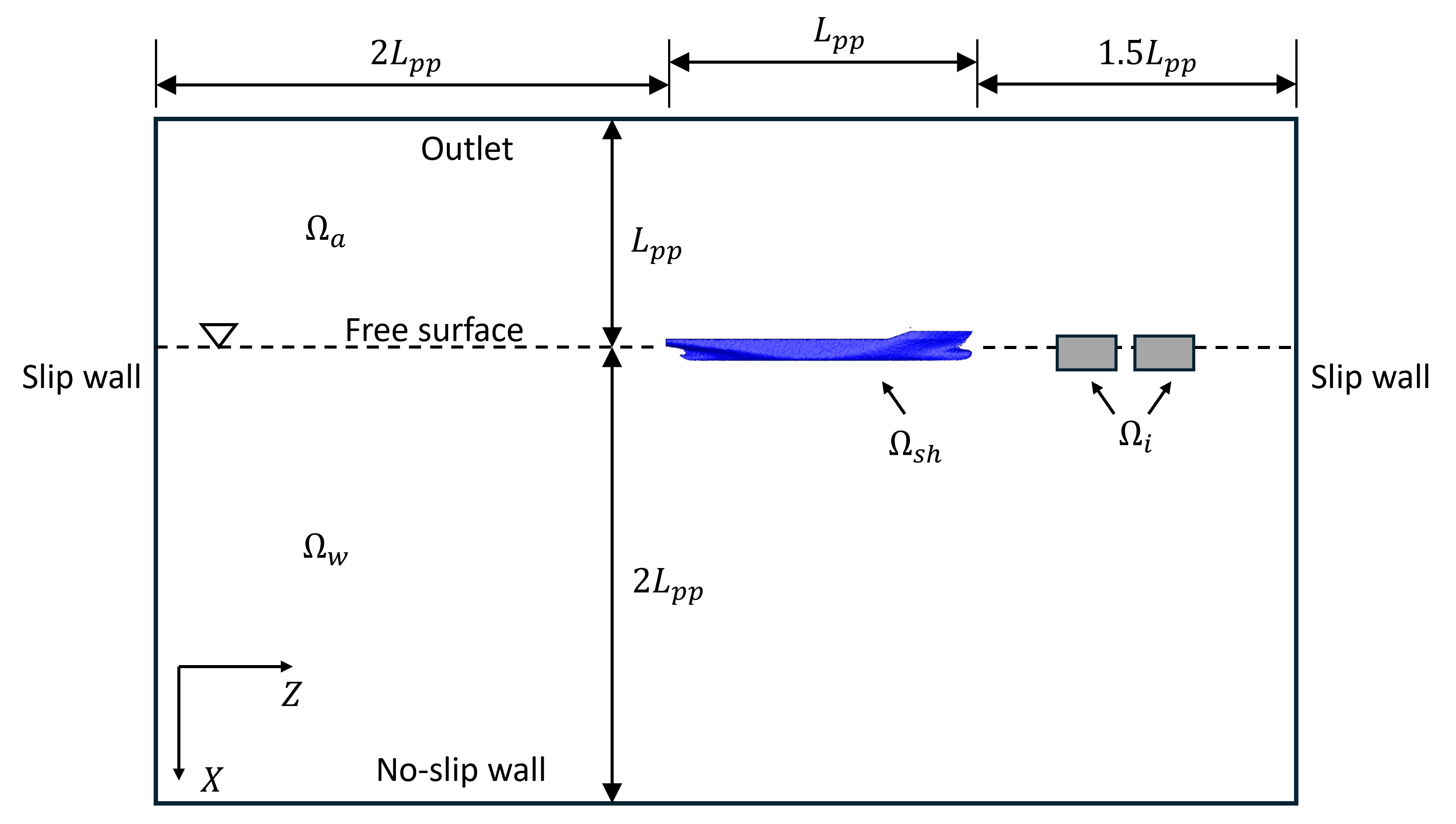}
        \caption{}
        \label{}
    \end{subfigure}
    \centering
    \caption{Ship-ice interaction problem: Schematic of the computational domain from (a) top view and (b) side view for the ship-ice interaction problem. $L_{pp}$ denotes the length between the perpendiculars of the ship.}
    \label{fig:si_schematic}
\end{figure}

\begin{figure}[htbp]
    \begin{subfigure}[h]{0.95\textwidth}
        \centering
        \includegraphics[width=1\textwidth]{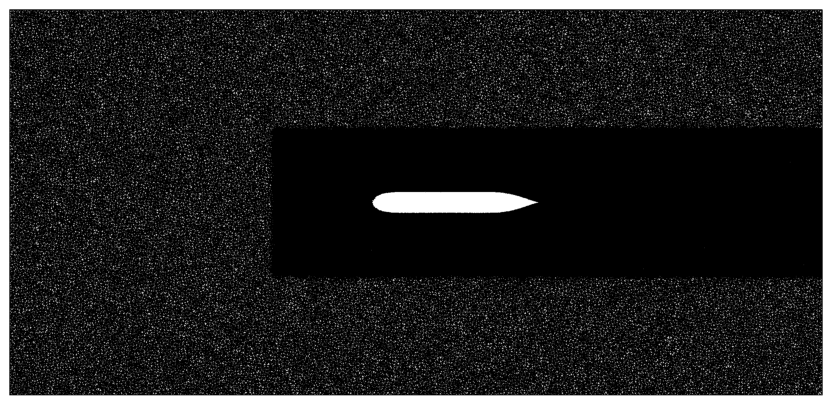}
        \caption{}
        \label{}
    \end{subfigure}
    \begin{subfigure}[h]{0.95\textwidth}
        \centering
        \includegraphics[width=1\textwidth]{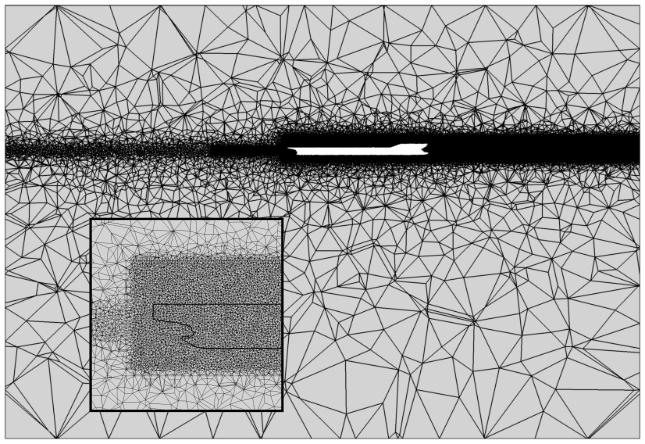}
        \caption{}
        \label{}
    \end{subfigure}
    \centering
    \caption{Ship-ice interaction problem: Discretization of the computational domain from (a) top view (at $x=-0.009$) and (b) side view (at $y=0$) for the ship-ice interaction problem. The mesh is refined at the free-surface and near the hull of the ship. The inset in the bottom image shows the mesh refinement around the ship hull.}
    \label{fig:si_mesh}
\end{figure}

One of the key parameters during the design of a ship is the length between the perpendiculars ($L_{pp}$), which serves as the representative length scale of the ship in the flow. The computational domain extends $4.5L_{pp}$ in the streamwise direction ($z$-axis), $2L_{pp}$ in the transverse direction ($y$-axis), and $3L_{pp}$ in the vertical direction ($x$-axis), as shown in Fig. \ref{fig:si_schematic}. These dimensions are in accordance with the ITTC guidelines \cite{procedures2014guidelines}. The origin of the domain is located at the stern of the ship. For the demonstration case, we employ the model-scale Kriso Container Ship (KCS), scaled $1:52.667$ from the full-scale vessel of length 230 m. The principal dimensions of the model ship are listed in Table~\ref{tab:KCS_dimensions}. The free surface is located at $x=-0.009$. Downstream, the ice field consists of two sets of pancake-model ice floes \cite{huang2020ship, sun2012simulation, thomson2018overview, bhuiyan2025assessment}, each with radius $R=0.4$ and thickness $t=0.3$.

\begin{table}
\caption{\label{tab:KCS_dimensions} Primary dimensions of the model scale KCS.}
\begin{center}
\begin{tabular}{c c} 
 \toprule
 Dimension & Value \\ [0.5ex] 
 \midrule
 Length between perpendiculars ($m$) & $4.367$ \\ 
 Waterline breadth ($m$) & $0.611$ \\ 
 Draft ($m$) & $0.2$ \\
 Wetted surface area ($m^2$) & $3.44$ \\
 Displacement ($m^3$) & $0.36$ \\[0.5ex] 
 \bottomrule
\end{tabular}
\end{center}
\end{table}

The Froude number, given by $Fr=\frac{v}{\sqrt{gL}}$, is an important non-dimensional parameter when performing hydrodynamic simulations with free-surface effects. We consider a Froude number of 0.12 for our simulation. The value of the assumed Froude number for our model scale is similar to that of typical ice-going ships operating in the Arctic. For a $g$ value of 9.81, we get the stream-wise velocity as $v_z=0.8$. 
This velocity is imposed by ramping it from zero within a small anchor zone inside the ship. The anchor zone is defined by the set of nodes satisfying $\sqrt{(\frac{x}{0.1})^2 + (\frac{y}{0.2})^2 + (\frac{z-d(t)}{1.8})^2} \le 1$, where $d(t)$ is the moving distance of the ship integrated from the imposed velocity $v_z$. The imposed velocity is given by the following definition
\begin{equation}
    v_z = \left\{
	\begin{array}{ll}
		v_0\sin{\left(\frac{2\pi t}{T}\right)}  & \mbox{if } t \leq T/4 \\
		v_0 & \mbox{if } t > T/4,
	\end{array}
\right.
\end{equation}
where $v_0=0.8$ and $T=10.5$. The other two components of velocity are set to zero, i.e. $v_y=v_x=0$, to avoid any transverse motion of the ship. 
The ice pieces are initialized sufficiently far downstream so that the ship enters the ice field after the initial acceleration phase. The top boundary is treated as an open boundary with zero-Neumann conditions. A no-slip condition is applied on the bottom wall, while slip-wall conditions are imposed on the remaining side walls. Zero-Neumann conditions are applied for the order-parameter fields on all boundaries. To determine a unique pressure field, $p=0$ is prescribed on the upper boundary.

\begin{figure}[htbp]
	\centering
	\begin{subfigure}{0.45\textwidth}
		\centering		
		\includegraphics[width=\textwidth]{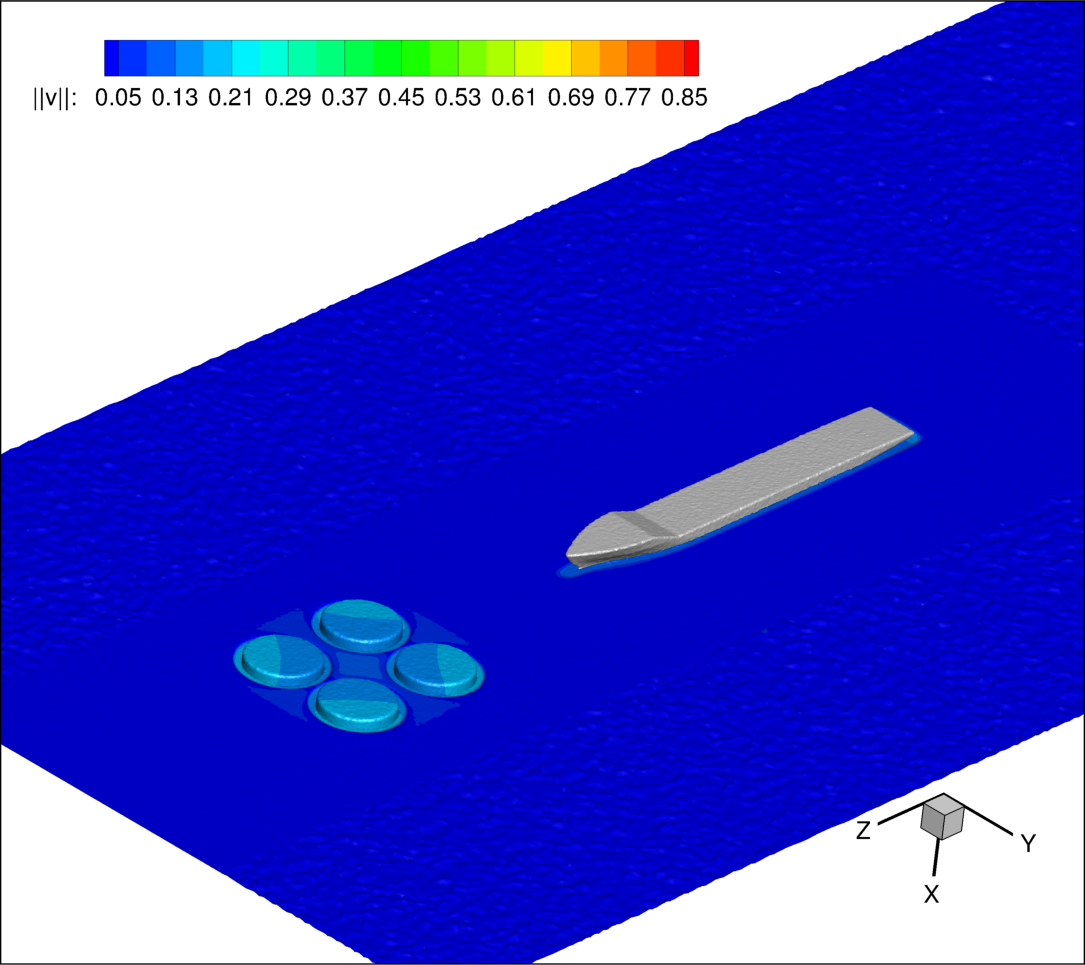}   
        \caption{}
	\end{subfigure}
	\hfill
	\begin{subfigure}{0.45\textwidth}
		\centering		
		\includegraphics[width=\textwidth]{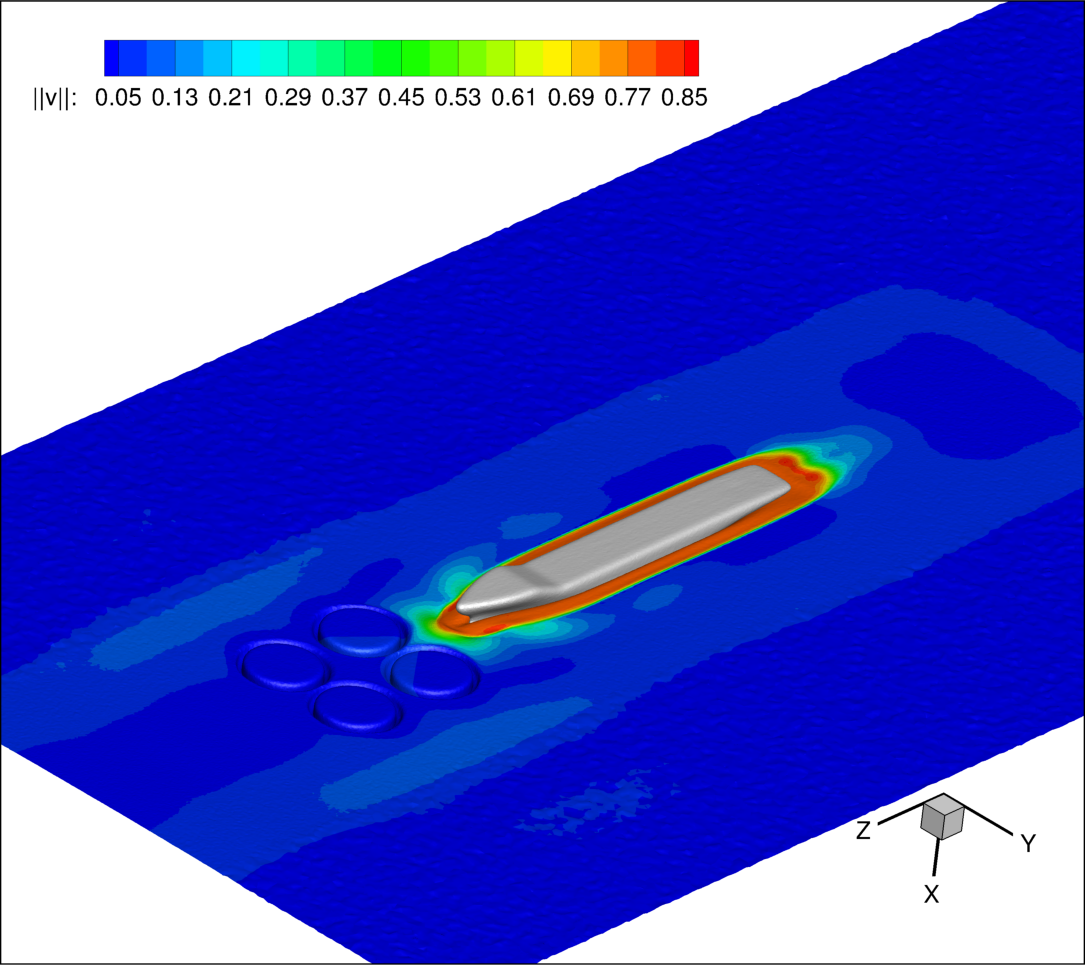}
		\caption{}
	\end{subfigure}
        \begin{subfigure}{0.45\textwidth}
		\centering		
		\includegraphics[width=\textwidth]{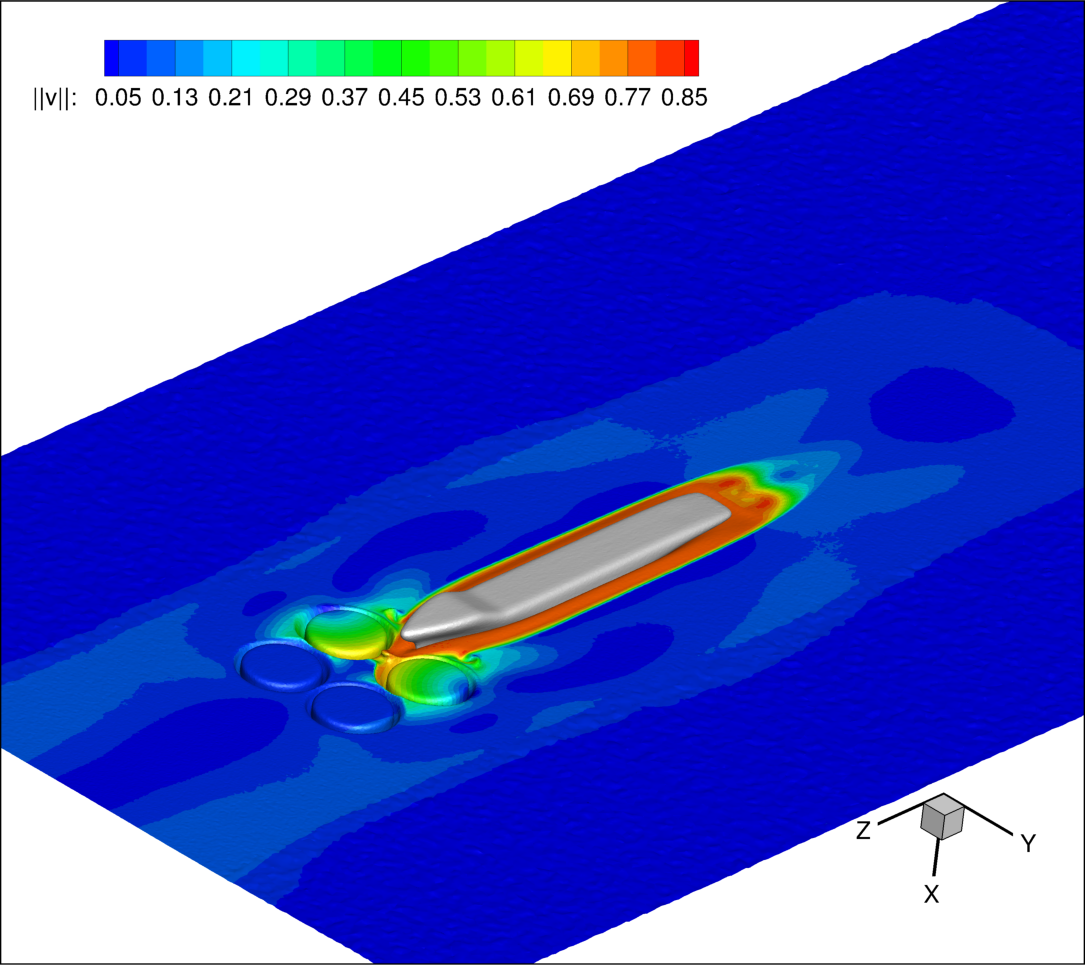}
		\caption{}
	\end{subfigure}
	\hfill
	\begin{subfigure}{0.45\textwidth}
		\centering		
		\includegraphics[width=\textwidth]{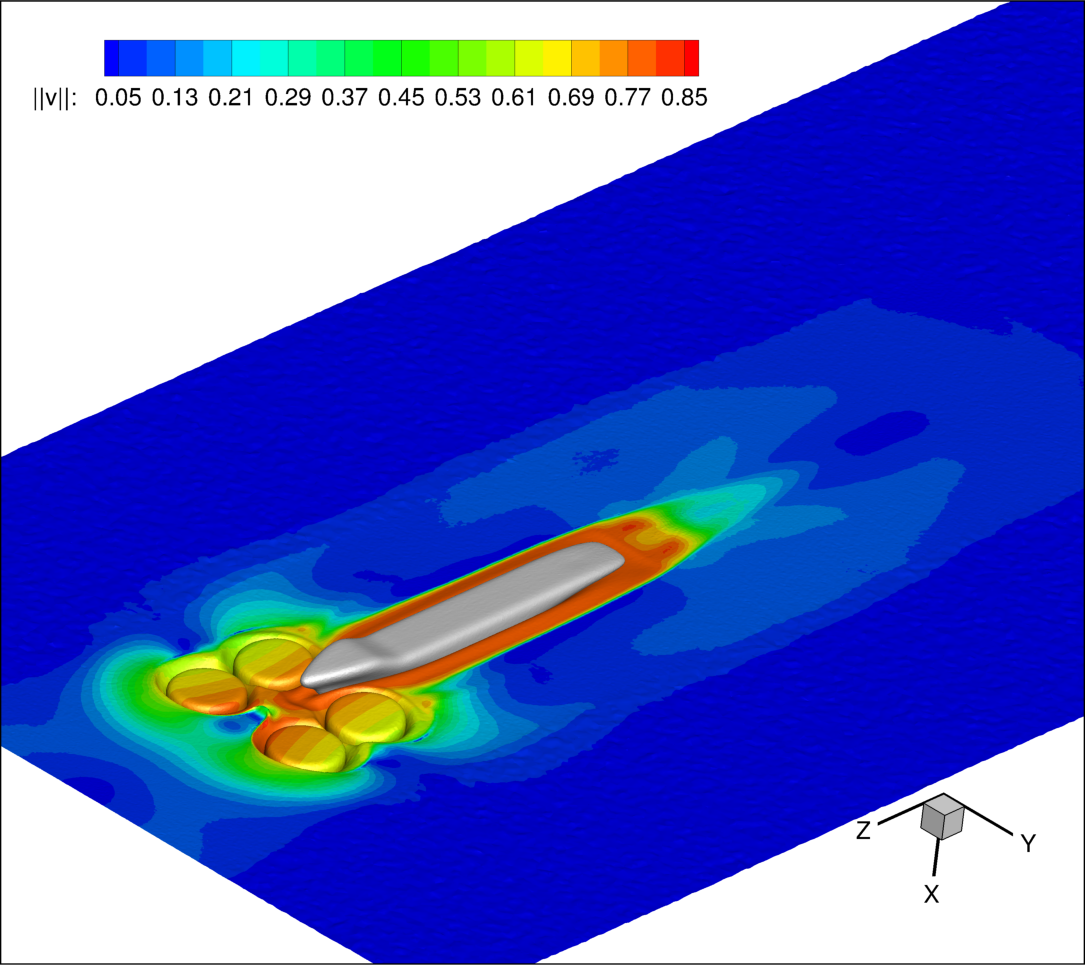}
		\caption{}
	\end{subfigure}
    \begin{subfigure}{0.45\textwidth}
		\centering		
		\includegraphics[width=\textwidth]{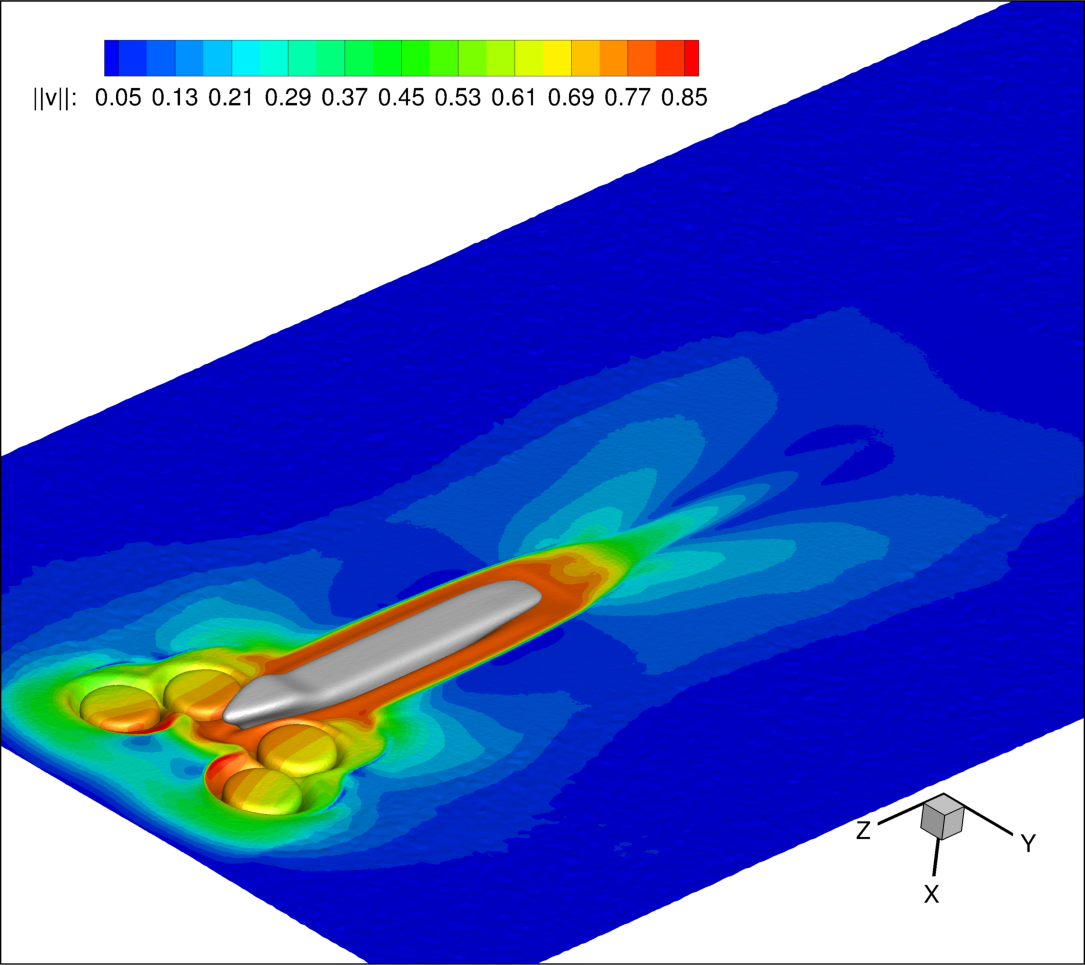}
		\caption{}
	\end{subfigure}
	\hfill
	\begin{subfigure}{0.45\textwidth}
		\centering		
		\includegraphics[width=\textwidth]{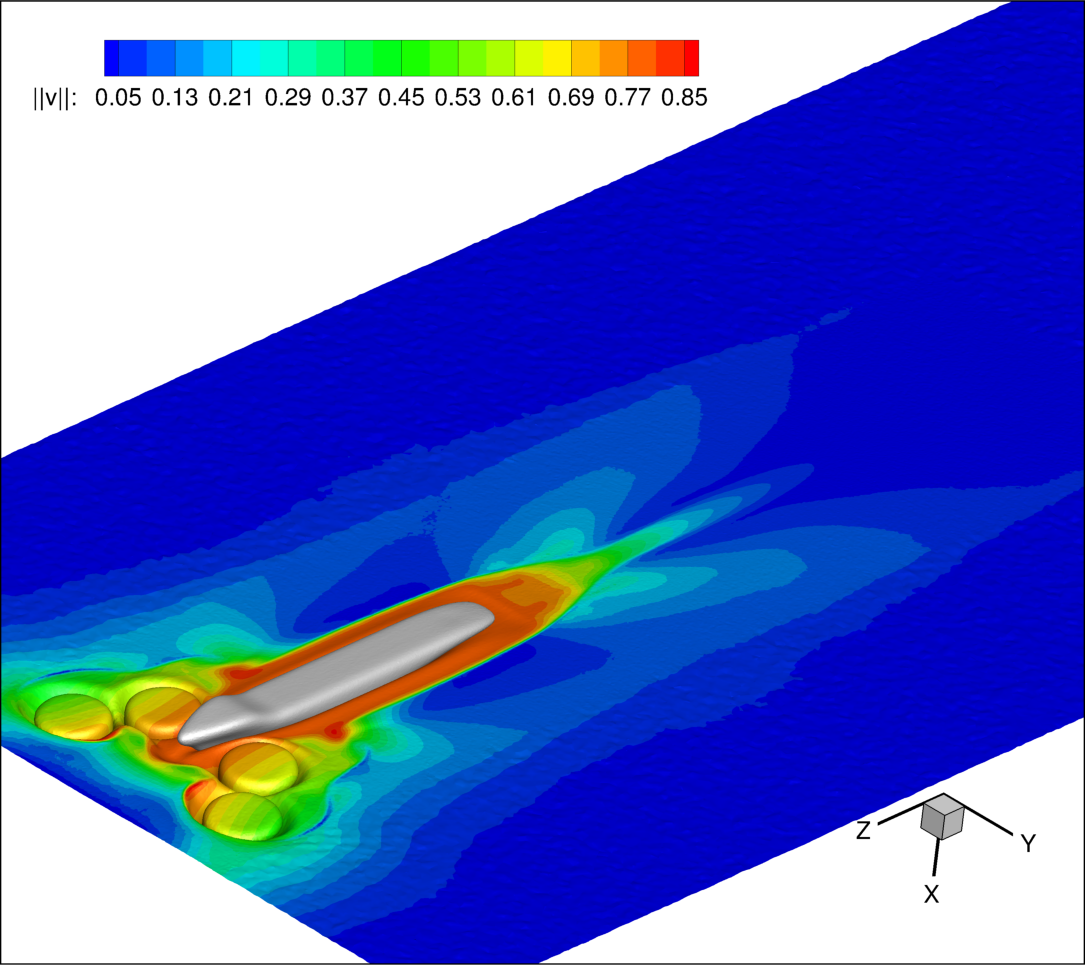}
		\caption{}
	\end{subfigure}
   
	\caption{Ship-ice interaction problem. Snapshots of the contours of magnitude of velocity at $t=$ (a) $0.2$, (b) $3$, (c) $4$, (d) $5.8$, (e) $7.2$, and (f) $8$. The iso-surfaces of $\phi=0$ for the ship, ice and water phases are plotted. The ship is represented using a solid gray colour.}
	\label{fig:si_plt_vMag}
\end{figure}

\begin{figure}[htbp]
	\centering
	\begin{subfigure}{0.45\textwidth}
		\centering		
		\includegraphics[width=\textwidth]{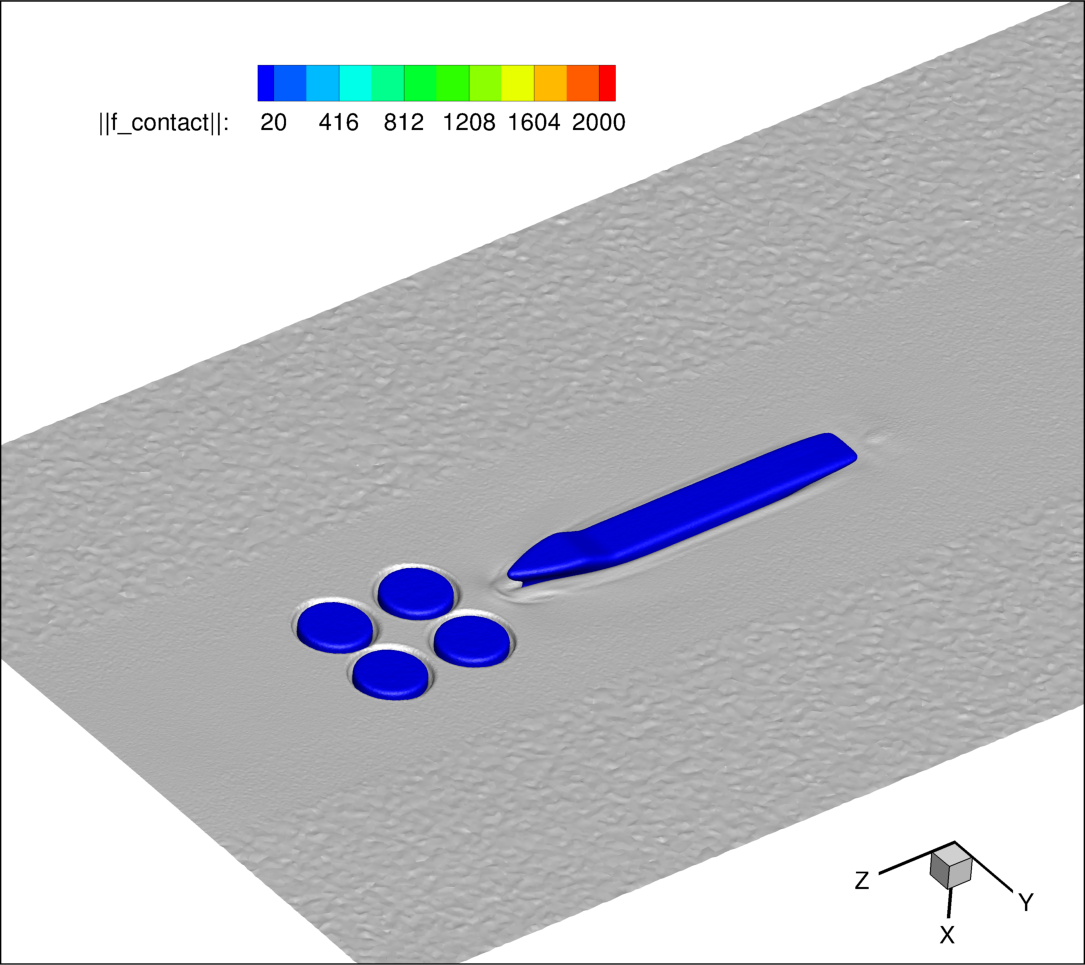}
        \caption{}
	\end{subfigure}
	\hfill
	\begin{subfigure}{0.45\textwidth}
		\centering		
		\includegraphics[width=\textwidth]{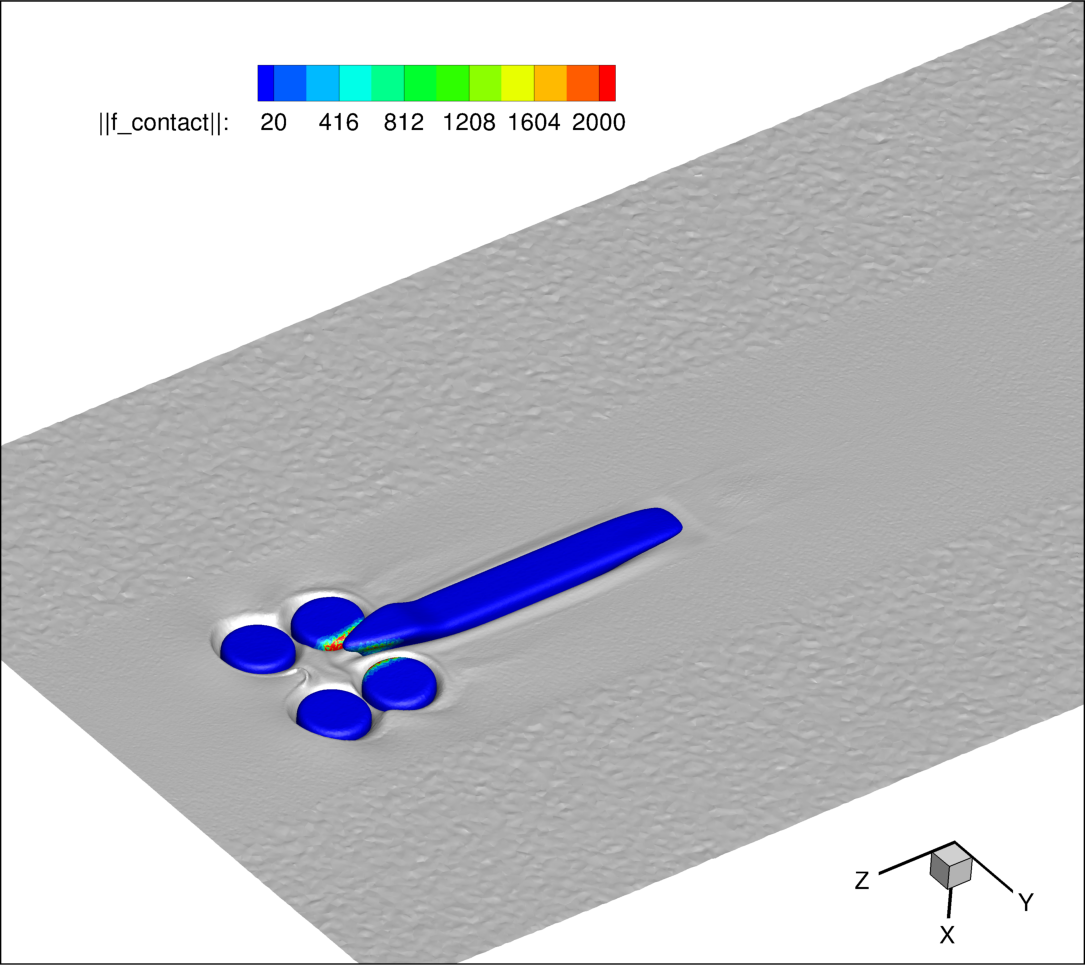}
		\caption{}
	\end{subfigure}
        \begin{subfigure}{0.45\textwidth}
		\centering		
		\includegraphics[width=\textwidth]{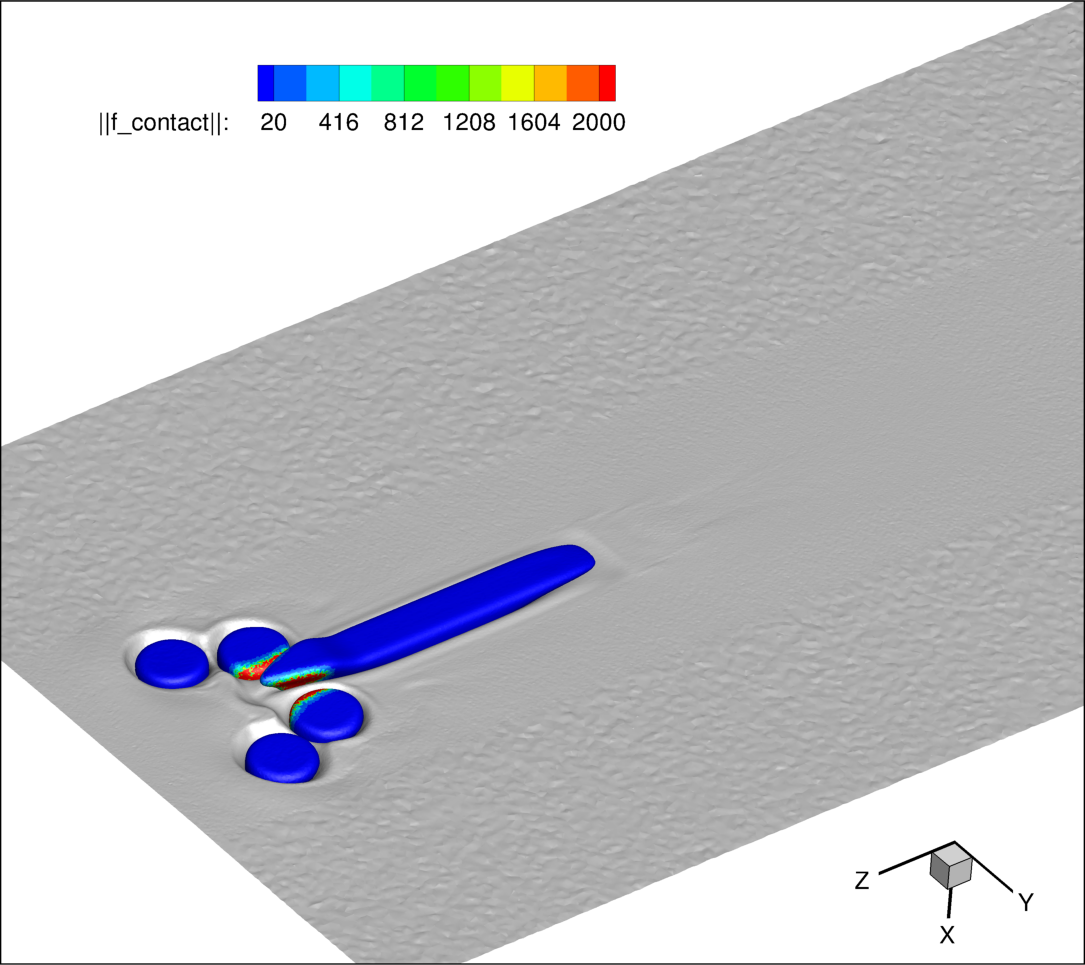}
		\caption{}
	\end{subfigure}
	\hfill
	\begin{subfigure}{0.45\textwidth}
		\centering		
		\includegraphics[width=\textwidth]{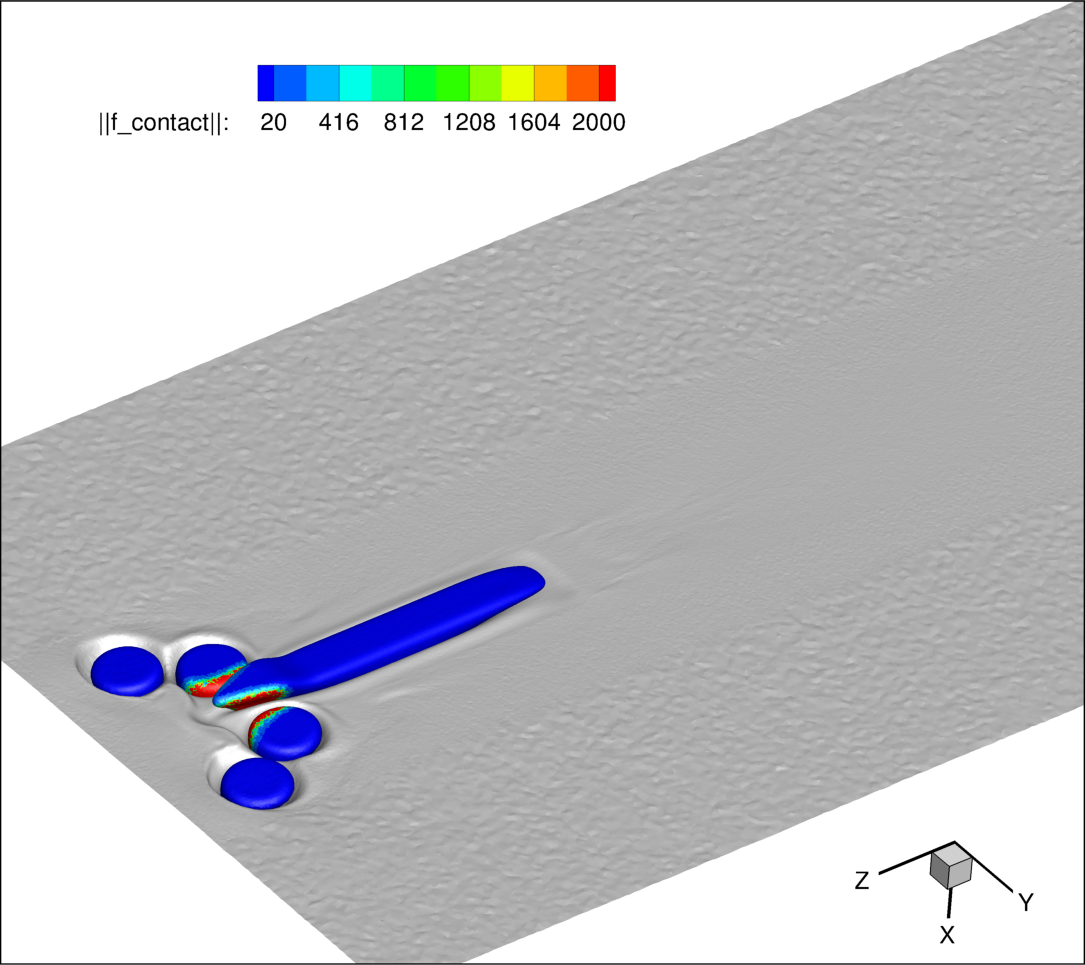}
		\caption{}
	\end{subfigure}   
	\caption{Ship-ice interaction problem. Snapshots of the contours of contact force magnitude at $t=$ (a) $3$, (b) $5.8$, (c) $7.2$, and (d) $8$ on the ship and ice floes. The iso-surfaces of $\phi=0$ for the ship, ice and water phases are plotted. The free-surface is represented using a solid gray colour.}
	\label{fig:si_plt_contactMag}
\end{figure}

The densities of the different phases are set as $\rho_{sh}=536$, $\rho_{i}=940$, $\rho_a=1.3$ and $\rho_w=1000$. The viscosities of the fluid phases are taken as $\mu_a=1.5 \times 10^{-5}$ and $\mu_w=10^{-3}$. Solid viscosities are assumed to be zero. The shear moduli for the solid phases are selected as $\mu_{sh}^L = 10^6$ and $\mu_i^L = 10^6$. The computational domain is discretized with tetrahedral elements of size $h=0.017$ around the interacting phases. It gradually coarsens away from the zone of interest, as shown in Fig. \ref{fig:si_mesh}. This results in an unstructured mesh of around $3.04 \times 10^6$ nodes for the discretized domain. The interface thickness parameter is chosen as $\varepsilon=h$ to have a reasonably thin diffuse interface without increasing the computational overhead. The contact parameters are chosen as $\kappa=10$ and $C_f=0.3$. The time step is selected as $\Delta t=0.01$ and the case is simulated until $t=8$. The problem was executed on 200 CPU cores and required approximately 17 hours to complete.

Figure~\ref{fig:si_plt_vMag} shows the results obtained from the 3D FSI-contact solver for this case. Contours of velocity magnitude illustrate the flow evolution around the interaction zone, with the free-surface wake clearly visible behind the ship. Figure~\ref{fig:si_plt_contactMag} illustrates the contours of the total contact forces acting on the ship and ice floes, highlighting the collision zones near the bow and shoulder regions of the ship. These plots demonstrate the capability of the unified phase-field model to capture the coupled dynamics of free-surface effects, FSI, and contact phenomena.

\begin{figure}[htbp]
    \centering
    \includegraphics[width=0.7\linewidth]{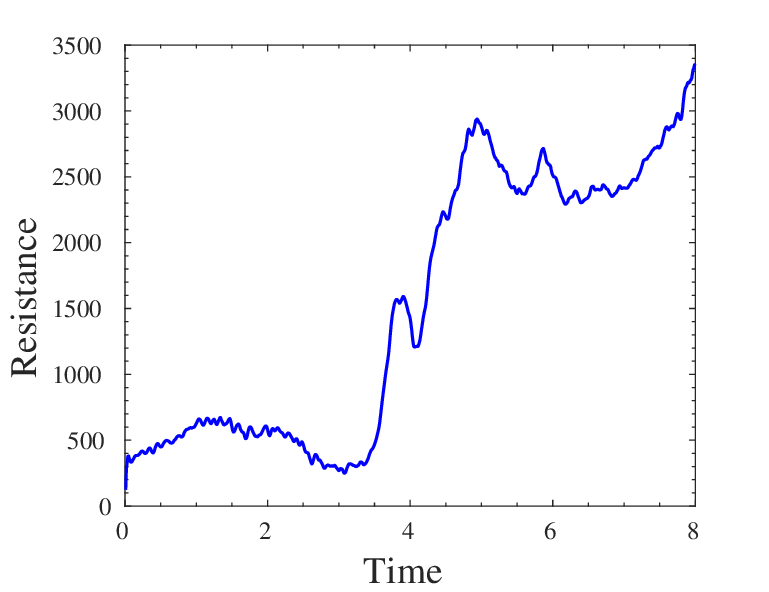}
    \caption{Ship-ice interaction problem. Temporal variation of the total resistance acting on the ship.}
    \label{fig:si_resistance}
\end{figure}

The total resistance, i.e., the force acting in the streamwise direction on the ship, is plotted in Fig.~\ref{fig:si_resistance}. The total force is obtained by integrating the stress divergence term within the ship, $\int_{\Omega_{sh}} \nabla \cdot \boldsymbol{\sigma}, d\Omega$.
The initial transient behavior, up to $t\approx2.63$, is attributed to the acceleration of the ship during this time. The ship resistance increases beyond $t=3.5$, when it encounters the ice-field. This increase is a combination of both novel hydrodynamic effects as well as normal and frictional contact of the ice floes with the ship. The peak at $t=5.8$ corresponds to the closest approach of the ship to the first row of ice blocks (Fig. \ref{fig:si_plt_vMag}). The curve dips temporarily due to the sliding of the ice pieces along the hull of the ship. The final increase in total resistance is due to the close proximity of the ship to the front wall. We stop the simulation at this point.

Using this simplified ship-ice interaction case, we have demonstrated that the proposed model can capture the complex dynamics of the problem while providing estimates of the total resistance acting on the ship. The diffuse-interface contact strategy exhibits robust performance in 3D settings, enabling extensions to industrial-scale marine scenarios. Further parametric studies and validation are required to assess the accuracy of the model for realistic predictions of ice-going ships. The developed framework can also be extended to capture damage mechanics of the ship hull and level ice by incorporating advanced ice mechanics formulations together with plasticity and failure models.

\section{Conclusions} \label{sec8}
In this article, we presented a phase-field based contact formulation for modeling frictional interactions in three dimensions. Building on our previously developed multiphase FSI framework, we introduced a novel approach to capture collisions between multi-component solids. The use of distinct phase-field functions enabled the definition of an overlap parameter to quantify proximity between colliding bodies, thereby bypassing explicit distance computations in 3D. The common normal was obtained from the gradients of the phase-field functions, while phase-averaged velocities were introduced to evaluate the relative sliding velocity. This formulation permits the use of a single momentum balance equation, reducing the number of degrees of freedom and maintaining consistency within the Eulerian framework.

The model was implemented on a hybrid MPI/OpenMP platform to simulate large-scale 3D problems in parallel. Verification against the Hertzian contact problem demonstrated an accurate prediction of contact traction profiles. The sliding block problem validated the dynamic frictional response, while the ironing problem highlighted robustness under finite deformation and large sliding motion.  Finally, we demonstrated the capabilities of the proposed framework by simulating a simplified ship-ice interaction problem by passing a representative container ship through a small ice-field consisting of pancake ice floes.

\section*{Acknowledgement} \label{sec:acknowledgement}
The authors would like to acknowledge the Natural Sciences and Engineering Research Council of Canada (NSERC) and Seaspan Shipyards for the funding. This research was supported in part through computational resources and services provided by Advanced Research Computing at the University of British Columbia.


\bibliography{ref.bib}

\end{document}